\documentclass[pra,twocolumn,tightenlines,nofootinbib,superscriptaddress]{revtex4}
\usepackage{graphicx} 
\usepackage{amsmath}
\usepackage{bm}
\usepackage{dcolumn}

\begin{document}

\title{Relativistic all-order many-body calculation of energies, wavelengths, and $M1$ and $E2$ transition
rates for  the $3d^n$ configurations in tungsten ions}

\author{M.~S.~Safronova}
\affiliation{Department of Physics and Astronomy, University of Delaware, Newark, DE 19716, USA}
\affiliation{Joint Quantum Institute, National Institute of Standards and Technology and the University of Maryland,
College Park, Maryland, 20742, USA}

\author{U.~I.~Safronova}
\affiliation{Department of Physics, University of Nevada, Reno, Nevada 89557, USA}

\author{S.~G.~Porsev}
\affiliation{Department of Physics and Astronomy, University of Delaware, Newark, DE 19716, USA}
\affiliation{Petersburg Nuclear Physics Institute, Gatchina, Leningrad District 188300, Russia}

\author{M. G. Kozlov}
\affiliation{Petersburg Nuclear Physics Institute, Gatchina, Leningrad District 188300, Russia}
\affiliation{St.~Petersburg Electrotechnical University ``LETI'', Prof. Popov Str. 5, 197376 St.~Petersburg}

\author{Yu. Ralchenko}
\affiliation{Atomic Spectroscopy Group, National Institute of Standards and Technology, Gaithersburg, MD 20899-8422, USA}

\date{\today}
\begin{abstract}

Energy levels, wavelengths,  magnetic-dipole and electric-quadrupole transition
rates between the low-lying states are evaluated for W$^{51+}$ to W$^{54+}$ ions
with $3d^n$ ($n$ = 2 to 5) electronic configurations using an approach combining
configuration interaction with linearized coupled-cluster single-double  method.
The QED corrections are directly incorporated into the calculations and their
effect is studied in detail. Uncertainties of the calculations are discussed.
This first study of such highly charged ions with the present method opens the
way for future applications allowing an accurate prediction of properties for a
very wide range of highly charged ions aimed at providing precision benchmarks
for various applications.

\end{abstract}
\maketitle

\section{Introduction}

Theoretical and experimental studies of tungsten highly charged ions  with an
open $3d$ shell is at present a subject of extensive
research~\cite{ralch-11,OsiGilRal12,ADNDT2017,atom-17}, motivated, in part, by
the proposed use of tungsten as a plasma-facing material in the divertor region
of the international reactor ITER \cite{HawCamJan09}. The core temperatures on
the order of 10--20 keV are not sufficient to completely ionize tungsten, and
its partially ionized atoms are expected to strongly emit in the X-ray and
extreme ultraviolet (EUV) ranges of spectra. The measured radiation can be
reliably used to diagnose certain plasma properties such as temperature and
density. This application stimulated an extensive analysis of the EUV spectra
between 10 and 25 nm from highly charged ions of tungsten with an open $3d$
shell~\cite{ralch-11} carried out at NIST. Using an electron-beam ion trap
(EBIT), a number of forbidden magnetic-dipole lines within ground configurations
of all $3d^n$ ions of tungsten, from Co-like W$^{47+}$ to K-like W$^{55+}$ were
measured and identified in the spectra~\cite{ralch-11}. This work demonstrated
that almost all strong lines were due to the forbidden magnetic-dipole ($M1$)
transitions within $3d^n$ ground configurations. Further study of
extreme-ultraviolet $M1$ lines in 50-60-fold ionized atoms of tungsten, hafnium,
tantalum, and gold with an open $3d$ shell was reported in~\cite{ralch-13}.
Using EBIT the spectra were measured at NIST and large-scale
collisional-radiative modeling was instrumental in line identification and in
analysis of their diagnostic potential. Furthemore, the $M1$ line ratios were
shown to be an accurate and versatile tool for studying the dielectronic
resonances in $3d^n$ ions including effects of anisotropy of the EBIT electron
energy distribution function \cite{M1DR}.

Motivated by such interest in the M1 transitions in open-shell highly charged
ions (HCIs) with $3d^n$ configuration, we  carry out a high-precision benchmark
study of tungsten HCIs using the state-of-the-art method which combined
configuration interaction (CI) with the linearized coupled-cluster method,
refereed to as the CI+all-order method \cite{SafKozJoh09}. The CI+all-order
method was used to accurately evaluate properties of atomic systems with two to
four valence electrons
\cite{SafKozSaf12,ZuhSafKoz12,SafPorKoz12,PorSafKoz12a,SafSafPor13,SafMaj13,SafDzuFla14,igor15},
including superheavy elements No, Lr and Rf~\cite{DzuSafSaf14}. An advantage of
this approach is an ability to include core correlations for all core shells
together with the accurate description of the valence electronic correlations.
The method was recently applied to low/medium ionization charge HCIs, up to
17$^+$ \cite{SafDzuFla14,SafDzuFla14a,SafDzuFla14b,DzuSafSaf15} to predict their
properties relevant to the new proposals to use HCIs for the development of
ultra-precise frequency standards and a search for variation of the
fine-structure constant \cite{BerDzuFla10}. In 2016, the QED corrections were
incorporated directly into the CI+all-order method by Tupitsyn {\it et al.\/}
\cite{qed-16}. The authors compared the performance of four different QED
potentials to estimate the accuracy of QED calculations and made a prediction of
HCI properties urgently needed for planning future experiments of interest to
metrology and tests of fundamental physics \cite{qed-16}.

In this work, we use a new version of the CI+all-order+QED method developed in
\cite{qed-16} to study, for the first time, the properties of HCIs with much
higher degree of ionization, up to W$^{54+}$. We also conducted a detailed study
of the QED corrections, carrying out all of the calculations with and without
inclusion of the QED to demonstrate the size of the QED contribution. This
effort paves the way for future applications of this approach for an accurate
prediction of properties for a very wide range of HCIs and providing precision
benchmarks for spectra identification and other applications. This is also the
first calculation of the system with five valence electrons with the
CI+all-order method.


We start with a summary of previous relevant theoretical studies. Energy levels
of the $3d^k$, $k = 1-9$, configurations for tungsten ions, computed using the
fully relativistic multiconfiguration Dirac-Hartree-Fock (MCDHF) GRASP2K code \cite{JonGaiBie13},
based on the variational method, were reported by Froese Fisher {\it et al.\/}
\cite{atom-17}. The correlation corrections for the $3s$, $3p$, $3d$ orbitals
considered to be valence orbitals, as well as the core-core and core-valence
effects from the $2s$, $2p$ subshells, were included in the calculations.
Extensive MCDHF calculations were also performed for the $3s^2 3p^6 3d^k$
($k=1-9$) ground configurations of HCIs with $Z=72-83$ in \cite{ADNDT2017}.
Complete and consistent data sets of excitation energies, wavelengths, line
strengths, oscillator strengths, and magnetic dipole and electric quadrupole
(E2) transition rates among all these levels were given and compared  with the
results available in the literature.

The wavelengths and transition probabilities were computed in~\cite{w-huang-15}
for forbidden transitions within the $3d^k$ ($k = 1-9$) ground configurations in
ions of hafnium, tantalum, tungsten, and gold. The authors used  the
second-order relativistic many-body perturbation theory (RMBPT) followed the
method described  in Ref.~\cite{lindgren-74}.

\begin{table*}[tbp]
\caption{\label{tab-en-2e} Ca-like W$^{54+}$. The  low-lying energy levels (in
cm$^{-1}$) calculated using the CI+all-order method are given in columns
``BREIT'' and ``QED''. They are compared with the recommended NIST
data~\cite{nist-web}, labeled as ``NIST'', and theoretical results from
Refs.~\cite{safr-ion-10} and \cite{atom-17}, labeled as ``RMBPT'' and
``GRASP2K'', respectively. First row gives the absolute value of the ground
state valence energy. The energies of the excited states are counted from the
ground state energy. The columns labeled ``BREIT'' and ``QED'' list the results
which include the Breit interaction  obtained without and with the QED
corrections, respectively. The differences between the ``NIST'' and ``RMBPT'',
``NIST'' and ``BREIT'', and ``NIST'' and ``QED'' values are shown in \% and
cm$^{-1}$ in columns labeled ``N$-$R'', ``N$-$B'', and ``N$-$Q'', respectively.}
\begin{ruledtabular}
\begin{tabular}{crrrrrrcrrrr}
\multicolumn{1}{c}{Level} &
\multicolumn{1}{r}{NIST~\cite{nist-web}}&
\multicolumn{1}{r}{{RMBPT~\cite{safr-ion-10}}} &
\multicolumn{1}{r}{{GRASP2K~\cite{atom-17}}} &
\multicolumn{1}{c}{{BREIT}} &
\multicolumn{1}{c}{{QED}} &
\multicolumn{3}{c}{Difference in \%}&
\multicolumn{3}{c}{Difference in cm$^{-1}$}\\
\multicolumn{6}{r}{} &
\multicolumn{1}{r}{N$-$R} &
\multicolumn{1}{r}{N$-$B} &
\multicolumn{1}{c}{N$-$Q} &
\multicolumn{1}{r}{N$-$R} &
\multicolumn{1}{r}{N$-$B} &
\multicolumn{1}{r}{N$-$Q} \\
\hline \\[-0.7pc]
$3d^2~ ^3\!F_2$& 85150000 &          &          & 85045940 & 85053834 &       &  0.12 &   0.11 &       & 104060 & 96166\\
$3d^2~ ^3\!P_0$&   188000 &   187110 &   186230 &   186228 &   186100 &  0.47 &  0.94 &   1.01 &   890 &   1772 &  1900\\
$3d^2~ ^3\!F_3$&   585480 &   582850 &   584750 &   584212 &   585659 &  0.45 &  0.22 &  -0.06 &  2630 &   1268 &  -379\\
$3d^2~ ^3\!D_2$&   668490 &   666210 &   667960 &   667321 &   668700 &  0.34 &  0.17 &  -0.03 &  2280 &   1169 &  -210\\
$3d^2~ ^3\!G_4$&   697000 &   693810 &   696100 &   695931 &   697355 &  0.46 &  0.15 &  -0.05 &  3190 &   1069 &  -355\\
$3d^2~ ^3\!P_1$&   709460 &   705410 &   706750 &   706048 &   707428 &  0.57 &  0.48 &   0.29 &  4050 &   3412 &  2032\\
$3d^2~ ^3\!F_4$&  1234000 &  1231640 &  1235570 &  1234504 &  1237339 &  0.19 & -0.04 &  -0.27 &  2360 &   -504 & -3339\\
$3d^2~ ^1\!D_2$&  1299000 &  1296730 &  1300180 &  1298669 &  1301477 &  0.17 &  0.03 &  -0.19 &  2270 &    331 & -2477\\
$3d^2~ ^1\!S_0$&  1493000 &  1491540 &  1493710 &  1492483 &  1495148 &  0.10 &  0.03 &  -0.14 &  1460 &    517 & -2148
\end{tabular}
\end{ruledtabular}
\end{table*}
\begin{table*}
\caption{\label{comp-2e-1} Wavelengths (in nm) and $A_{M1}$ and $A_{E2}$ transition rates (in s$^{-1}$) in W$^{54+}$
are compared with the available theoretical~\cite{safr-ion-10} and experimental~\cite{ralch-11} results.
Numbers in brackets represent powers of 10.}
\begin{ruledtabular}
\begin{tabular}{lrrrrrrrrrrrrr}
\multicolumn{2}{c}{Transition} &
\multicolumn{3}{c}{$\lambda$, nm} &
\multicolumn{2}{c}{$A_{M1}$} &
\multicolumn{2}{c}{$A_{E2}$} \\
\multicolumn{1}{l}{final}&
\multicolumn{1}{r}{initial}&
\multicolumn{1}{c}{Present} &
\multicolumn{1}{c}{Ref.~\cite{safr-ion-10}} &
\multicolumn{1}{c}{Ref.~\cite{ralch-11}} &
\multicolumn{1}{c}{Present} &
\multicolumn{1}{c}{Ref.~\cite{safr-ion-10}} &
\multicolumn{1}{c}{Present} &
\multicolumn{1}{c}{Ref.~\cite{safr-ion-10}} \\
\hline \\ [-0.7pc]
$^3\!F_2$ & $^1\!D_2$ &   7.700  &   7.712 &        &  1.14[4]  & 1.276[4] &  1.26[1] &  4.507[1]\\
$^3\!F_2$ & $^3\!F_4$ &   8.100  &   8.119 &        &           &          &  2.21[2] &  4.304[2]\\
$^3\!P_1$ & $^1\!S_0$ &  12.716  &  12.721 &        &  7.79[6]  & 7.323[6] &          &          \\
$^3\!F_3$ & $^1\!D_2$ &  13.997  &  14.008 &        &  7.48[5]  & 7.524[5] &  1.06[3] &  1.052[3]\\
$^3\!F_2$ & $^3\!P_1$ &  14.163  &  14.176 &        &  2.60[5]  & 2.583[5] &  9.67[2] &  1.179[3]\\
$^3\!F_2$ & $^3\!G_4$ &  14.369  &  14.413 &        &           &          &  4.22[2] &  3.219[2]\\[0.4pc]

$^3\!F_2$ & $^3\!D_2$ &  14.985  &  15.010 & 14.959 &  1.79[6]  & 1.798[6] &  7.74[2] &  7.312[2]\\
$^3\!F_3$ & $^3\!F_4$ &  15.378  &  15.413 &        &  3.79[6]  & 3.755[6] &  7.19[1] &  6.133[1]\\
$^3\!D_2$ & $^1\!D_2$ &  15.839  &  15.860 &        &  3.08[6]  & 3.095[6] &  1.00[2] &  7.536[1]\\
$^3\!G_4$ & $^1\!D_2$ &  16.591  &  16.586 &        &           &          &  1.41[0] &  1.998[1]\\
$^3\!P_1$ & $^1\!D_2$ &  16.874  &  16.911 &        &  1.29[6]  & 1.285[6] &  3.56[2] &  4.184[2]\\[0.4pc]
$^3\!F_2$ & $^3\!F_3$ &  17.117  &  17.157 & 17.080 &  3.66[6]  & 3.683[6] &  1.23[2] &  1.154[2]\\

$^3\!D_2$ & $^3\!F_4$ &  17.631  &  17.686 &        &           &          &  1.84[2] &  6.397[1]\\
$^3\!G_4$ & $^3\!F_4$ &  18.568  &  18.593 &        &  1.09[6]  & 1.110[6] &  3.51[2] &  7.548[2]\\[0.4pc]
$^3\!P_0$ & $^3\!P_1$ &  19.237  &  19.294 & 19.177 &  1.72[6]  & 1.771[6] &          &            \\

$^3\!F_3$ & $^3\!P_1$ &  82.078  &  81.595 &        &           &          &  8.20[-1]&  9.071[-1]\\
$^3\!F_3$ & $^3\!G_4$ &  89.510  &  90.123 &        &  8.41[3]  & 8.556[3] &  2.70[-3]&  4.237[-5]\\
$^3\!F_3$ & $^3\!D_2$ & 120.325  & 119.974 &        &  4.32[3]  & 4.351[3] &  1.14[-2]&  2.145[-2]\\
$^3\!F_4$ & $^1\!D_2$ & 155.851  & 153.641 &        &           &          &  2.78[-2]&  3.740[-3]\\
$^3\!D_2$ & $^3\!P_1$ & 258.211  & 253.066 &        &  6.40[2]  & 6.788[2] &  2.88[-3]&  3.731[-3]\\
$^3\!D_2$ & $^3\!G_4$ & 349.516  & 362.222 &        &           &          &  1.90[-4]&  8.126[-4]\\
\end{tabular}
\end{ruledtabular}
\end{table*}

\begin{table}
\caption{\label{comp-2e-2} Wavelengths (nm) for transitions within the $3d^2$
configuration in Ca-like W$^{54+}$  evaluated using the CI+all-order method without and with QED contributions.
The wavelengths are compared with available measurements from Ref.~\cite{ralch-11}.}
\begin{ruledtabular}
\begin{tabular}{lrrrr}
\multicolumn{2}{c}{Transition} &
\multicolumn{3}{c}{Wavelengths, $\lambda$, nm} \\
\multicolumn{2}{c}{}&
\multicolumn{1}{c}{NoQED}&
\multicolumn{1}{c}{QED} &
\multicolumn{1}{c}{Expt.~\cite{ralch-11}} \\
\hline \\ [-0.7pc]
   $3d^2\  ^3\!F_{2} $&$3d^2\ ^1\!S_{0}$&     6.700 &     6.688&\\
   $3d^2\  ^3\!F_{2} $&$3d^2\ ^1\!D_{2}$&     7.699 &     7.684&\\
   $3d^2\  ^3\!F_{2} $&$3d^2\ ^3\!F_{4}$&     8.099 &     8.082&\\
   $3d^2\  ^3\!P_{0} $&$3d^2\ ^1\!D_{2}$&     8.987 &     8.966&\\
   $3d^2\  ^3\!P_{2} $&$3d^2\ ^1\!S_{0}$&    12.119 &    12.100&\\
   $3d^2\  ^3\!P_{1} $&$3d^2\ ^1\!S_{0}$&    12.716 &    12.695&\\
   $3d^2\  ^3\!F_{3} $&$3d^2\ ^1\!D_{2}$&    13.996 &    13.970&\\
   $3d^2\  ^3\!F_{2} $&$3d^2\ ^3\!P_{1}$&    14.162 &    14.136&\\
   $3d^2\  ^3\!F_{2} $&$3d^2\ ^1\!G_{4}$&    14.368 &    14.340&\\ $3d^2\ ^3\!F_{2} $&$3d^2\ ^3\!P_{2} $&  14.984 & 14.954&14.959\\
   $3d^2\  ^3\!F_{3} $&$3d^2\ ^3\!F_{4}$&    15.376 &    15.345&\\
   $3d^2\  ^3\!P_{2} $&$3d^2\ ^1\!D_{2}$&    15.837 &    15.803&\\
   $3d^2\  ^1\!G_{4} $&$3d^2\ ^1\!D_{2}$&    16.588 &    16.553&\\
   $3d^2\  ^3\!P_{1} $&$3d^2\ ^1\!D_{2}$&    16.871 &    16.834&\\ $3d^2\ ^3\!F_{2} $&$ 3d^2\ ^3\!F_{3} $& 17.113 & 17.075&17.080\\
   $3d^2\  ^3\!P_{2} $&$3d^2\ ^3\!F_{4}$&    17.627 &    17.586&\\
   $3d^2\  ^1\!G_{4} $&$3d^2\ ^3\!F_{4}$&    18.563 &    18.519&\\ $3d^2\ ^3\!P_{0} $&$ 3d^2\ ^3\!P_{1} $& 19.230 & 19.182&19.177\\
   $3d^2\  ^3\!P_{0} $&$3d^2\ ^3\!P_{2}$&    20.777 &    20.721&\\
   $3d^2\  ^1\!D_{2} $&$3d^2\ ^1\!S_{0}$&    51.625 &    51.634&\\
   $3d^2\  ^3\!F_{2} $&$3d^2\ ^3\!P_{0}$&    53.740 &    53.735&\\
   $3d^2\  ^3\!F_{3} $&$3d^2\ ^3\!P_{1}$&    82.116 &    82.123&\\
   $3d^2\  ^3\!F_{3} $&$3d^2\ ^1\!G_{4}$&    89.544 &    89.529&\\
   $3d^2\  ^3\!F_{3} $&$3d^2\ ^3\!P_{2}$&   120.402 &   120.422&\\
   $3d^2\  ^3\!F_{4} $&$3d^2\ ^1\!D_{2}$&   155.926 &   155.914&\\
   $3d^2\  ^3\!P_{2} $&$3d^2\ ^3\!P_{1}$&   258.238 &   258.211&\\
   $3d^2\  ^3\!P_{2} $&$3d^2\ ^1\!G_{4}$&   349.382 &   348.979&\\
\end{tabular}
 \end{ruledtabular}
\end{table}
The excitation energies and transition rates for the states within the $3d^2$ configuration of
Ca-like ions with $Z$ = 22-100 were calculated by Safronova  {\it et al.\/} in Ref.~\cite{safr-3d2-01}.
The method based on RMBPT, including the Breit interaction, was used to evaluate the matrix
elements of $M1$ and $E2$ operators, including the retardation and contribution from
negative--energy states.
The wavelengths and $M1$ and $E2$ transition rates for Ca-like tungsten were
reported in~\cite{safr-ion-10}. The results were obtained in the framework of the RMBPT.
The first-order perturbation theory was used to obtain intermediate coupling coefficients and
the second-order RMBPT was used to determine the matrix elements.

The wavelengths and transition rates were computed by Quinet~\cite{quinet-11}
for forbidden transitions within the $3d^k$ ground configurations of tungsten
ions from W$^{47+}$ to W$^{61+}$ using  a fully relativistic multiconfiguration
Dirac-Fock method. The single and double excitations within the $n$ = 3 complex,
some $n = 3 \rightarrow n' = 4$ single excitations, the Breit interaction, and
the QED effects were included.

The atomic structure and spectra of ten tungsten ions were calculated using the
Flexible Atomic Code (FAC) by Clementson {\it et al.\/}~\cite{adndt-14}. The
energy levels, radiative lifetimes, spectral line positions, transition
probability rates, and oscillator strengths for the tungsten ions isoelectronic
to germanium, W$^{42+}$, through vanadium, W$^{51+}$, were reported.

\begin{table*}[tbp]
\caption{\label{tab-en-3e} The calculated energy levels of Sc-like W$^{53+}$ ion (in cm$^{-1}$) within the $3d^3$ configuration
are listed in the columns ``BREIT'' and ``QED''. They are compared with the recommended NIST data~\cite{nist-web} and theoretical results from Refs.~\cite{w-huang-15,atom-17}. First row gives the first ionization potential.
The excited state energies are counted from the ground state energy.
The differences (in \%) between ``FAC'' and ``NIST'', ``BREIT'' and ``NIST'', and ``QED'' and ``NIST'' values are presented by three last columns and labeled as ``F$-$N'', ``B$-$N'', and ``Q$-$N'', respectively.
The $g$ factors given by the Land\'{e} formula (``nr'') and calculated by the CI+all-order method (``BREIT'')
are presented in columns 7 and 8.}
\begin{ruledtabular}
\begin{tabular}{lrrrrrrrrrrr}
\multicolumn{1}{c}{Level} &
\multicolumn{5}{c}{Energies} &
\multicolumn{2}{c}{g-factor} &
\multicolumn{3}{c}{Difference (in \%)}\\
\multicolumn{1}{c}{} &
\multicolumn{1}{c}{NIST~\cite{nist-web}} &
\multicolumn{1}{c}{FAC~\cite{w-huang-15}} &
\multicolumn{1}{r}{{GRASP2K~\cite{atom-17}}} &
\multicolumn{1}{c}{BREIT} &
\multicolumn{1}{c}{QED} &
\multicolumn{1}{c}{nr} &
\multicolumn{1}{c}{BREIT} &
\multicolumn{1}{c}{F$-$N} &
\multicolumn{1}{c}{B$-$N} &
\multicolumn{1}{c}{Q$-$N} \\
\hline
$^2D^{1}_{3/2}$ &  40833000&           &          &   40770300&   40774500&  0.8000  &0.7211  &      &  -0.15 & -0.14      \\
$^4F^{1}_{5/2}$ &    530030&     530511&   529070 &     528021&     529544&  1.0286  &1.0512  & 0.09 &  -0.38  & -0.09  \\
$^4D^{1}_{3/2}$ &    580860&     580864&   579990 &     578123&     579594&  1.2000  &1.1659  & 0.00 &  -0.47  & -0.22  \\
$^4H^{ }_{9/2}$ &    610000&     611618&   610860 &     610047&     611577&  0.9697  &1.0321  & 0.26 &   0.01  &  0.26  \\
$^4G^{ }_{7/2}$ &    610000&     611860&   610320 &     610264&     611802&  0.9841  &1.0355  & 0.30 &   0.04  &  0.29 \\
$^2D^{1}_{5/2}$ &    812200&     812220&   812070 &     811936&     813338&  1.2000  &1.2363  & 0.00 &  -0.03  &  0.14  \\
$^2F^{1}_{7/2}$ &   1125950&    1126000&  1128600 &    1126911&    1129889&  1.1429  &1.1156  & 0.00 &   0.09  &  0.35  \\
$^2G^{ }_{9/2}$ &   1163858&    1164000&  1165990 &    1164377&    1167384&  1.1111  &1.1285  & 0.01 &   0.04  &  0.30  \\
$^4D^{2}_{3/2}$ &   1205798&    1206000&  1207730 &    1206422&    1209376&  1.2000  &1.1057  & 0.02 &   0.05  &  0.30  \\
$^2H^{ }_{11/2}$&   1243513&    1243000&  1243300 &    1242189&    1245195&  1.0909  &1.0791  &-0.04 &  -0.11  &  0.14  \\
$^2D^{2}_{5/2}$ &   1243706&    1244000&  1244610 &    1243569&    1246465&  1.2000  &1.2878  & 0.02 &  -0.01  &  0.22  \\
$^4F^{2}_{5/2}$ &   1314683&    1315000&  1315540 &    1314506&    1317477&  1.0286  &1.0911  & 0.02 &  -0.01  &  0.21  \\
$^2F^{2}_{7/2}$ &   1320329&    1320000&  1319550 &    1318007&    1320942&  1.1429  &1.0836  &-0.02 &  -0.18  &  0.05  \\
$^2D^{2}_{3/2}$ &   1481640&    1482000&  1481260 &    1480471&    1483413&  0.8000  &0.8856  & 0.02 &  -0.08  &  0.12  \\
$^4G^{ }_{9/2}$ &          &    1767023&  1764860 &    1762772&    1767221&  1.1717  &1.1576  &      &         &      \\
$^4D^{3}_{3/2}$ &          &    1878537&  1878320 &    1875314&    1879683&  1.2000  &1.1292  &      &         &      \\
$^2D^{3}_{5/2}$ &          &    1959564&  1960120 &    1958695&    1963041&  1.2000  &1.1607  &      &         &
\end{tabular}
\end{ruledtabular}
\end{table*}
\begin{table*}[tbp]
\caption{\label{comp-3el} Sc-like W$^{53+}$. Wavelengths (in nm) and $M1$ transition rates (in s$^{-1}$)
for the states within the $3d^3$ configuration are compared with the NIST data~\cite{ralch-11} where available.
Numbers in brackets represent powers of 10.}
\begin{ruledtabular}
\begin{tabular}{lrrrrrrr}
\multicolumn{2}{c}{Transition} &
\multicolumn{3}{c}{Wavelength (nm)}&
\multicolumn{3}{c}{$M1$ transition rate}\\
\multicolumn{1}{c}{final}&
\multicolumn{1}{c}{initial}&
\multicolumn{1}{c}{BREIT}&
\multicolumn{1}{c}{QED}&
\multicolumn{1}{c}{Ref.~\cite{ralch-11}}&
\multicolumn{1}{c}{BREIT}&
\multicolumn{1}{c}{QED}&
\multicolumn{1}{c}{Ref.~\cite{ralch-11}}\\
\hline \\[-0.6pc]
$^4F^1_{5/2} $&$^4D^2_{3/2}$ & 10.499  &   10.484 &       & 6.73[5] & 6.75[5] & \\
$^4D^1_{3/2} $&$^4D^2_{3/2}$ & 11.082  &   11.064 &       & 3.55[5] & 3.55[5] & \\[0.4pc]

$^2D^1_{3/2} $&$^2D^1_{5/2} $& 12.316  &   12.294 & 12.312& 2.78[5] &  2.80[5] & 2.75[5]\\

$^4F^1_{5/2} $&$^2F^2_{7/2} $& 12.658  &   12.636 &       & 1.60[5] &  1.60[5] &\\
$^4F^1_{5/2} $&$^2D^2_{5/2} $& 13.975  &   13.949 &       & 4.32[5] &  4.35[5] &\\
$^4H_{9/2}   $&$^2F^2_{7/2} $& 14.125  &   14.097 &       & 6.66[5] &  6.70[5] &\\
$^4G_{7/2}   $&$^2F^2_{7/2} $& 14.129  &   14.102 &       & 7.51[5] &  7.56[5] &\\
$^4G_{7/2}   $&$^4F^2_{5/2} $& 14.200  &   14.171 &       & 1.11[6] &  1.12[6] & \\
$^4F^1_{5/2} $&$^4D^2_{3/2} $& 14.741  &   14.710 &       & 1.97[6] &  1.98[6] & \\
$^2D^1_{5/2} $&$^4D^2_{3/2} $& 14.958  &   14.925 &       & 2.38[6] &  2.40[6] & \\
$^4D^1_{3/2} $&$^2D^2_{5/2} $& 15.028  &   14.995 &       & 2.48[6] &  2.50[6] & \\
$^4G_{7/2}   $&$^2D^2_{5/2} $& 15.790  &   15.756 &       & 2.02[5] &  2.02[5] &\\[0.4pc]
$^4H_{9/2}   $&$^2H_{11/2}  $& 15.819  &   15.782 & 15.785& 1.42[6] &  1.43[6] &1.42[6]\\
$^4F^1_{5/2} $&$^2F^1_{7/2} $& 16.698  &   16.657 &       & 4.90[6] &  4.94[6] &\\[0.4pc]
$^2D^1_{3/2} $&$^4D^1_{3/2} $& 17.297  &   17.253 & 17.216& 2.75[6] &  2.78[6] & 2.74[6]\\

$^4H_{9/2}   $&$^2G_{9/2}   $& 18.040  &   17.992 &       & 2.29[6] &  2.30[6] & \\
$^4G_{7/2}   $&$^2G_{9/2}   $& 18.047  &   17.999 &       & 1.30[6] &  1.31[6] &\\[0.4pc]
$^2D^1_{3/2} $&$^4F^1_{5/2} $& 18.939  &   18.884 & 18.867& 3.42[6] &  3.42[6] &3.41[6]\\

$^4G_{7/2}   $&$^2F^1_{7/2} $& 19.356  &   19.302 &       & 1.19[6] &  1.20[6] & \\
$^2D^1_{5/2} $&$^2F^2_{7/2} $& 19.760  &   19.702 &       & 2.54[5] &  2.56[5] &\\
$^2D^1_{5/2} $&$^4F^2_{5/2} $& 19.898  &   19.838 &       & 1.13[6] &  1.14[6] & \\
\end{tabular}
\end{ruledtabular}
\end{table*}

\begin{table*}
\caption{\label{tab-en-4ea} Ti-like W$^{52+}$ ion. The energies (in cm$^{-1}$) obtained in this work are listed in the
columns ``BREIT'' and ``QED''. They are compared with the recommended NIST data~\cite{nist-web} and theoretical results~\cite{w-huang-15,atom-17}. First row gives the first ionization potential. The energies of the excited states are counted
from the ground state energy. The nonrelativistic $g$ factor (``nr'') and $g$ factor obtained using the CI+all-order method
(``BREIT'') are presented.}
\begin{ruledtabular}
\begin{tabular}{lrrrrrrrr}
\multicolumn{1}{c}{Level} &
\multicolumn{5}{c}{Energy (cm$^{-1}$)} &
\multicolumn{2}{c}{g-factor} \\
\multicolumn{1}{c}{} &
\multicolumn{1}{c}{{NIST\cite{nist-web}}} &
\multicolumn{1}{c}{{FAC \cite{w-huang-15}}} &
\multicolumn{1}{r}{{GRASP2K~\cite{atom-17}}} &
\multicolumn{1}{c}{{BREIT}} &
\multicolumn{1}{c}{{QED}} &
\multicolumn{1}{c}{nr} &
\multicolumn{1}{c}{BREIT} \\
\hline \\[-0.6pc]
$ 3d^4\  ^3\!P^1_{0}$&   39739000&           &         &  39675100&   39679450&            &        \\
$ 3d^4\  ^3\!P_{1}$  &     517630&     518082&   516510&    514599&     516181&      1.5000&  1.4884\\
$ 3d^4\  ^3\!G^1_{4}$&     613000&     614788&   613540&    613414&     615007&      1.0500&  1.0604\\
$ 3d^4\  ^3\!D^1_{2}$&     638000&     639339&   638390&    636589&     638203&      1.1667&  1.1954\\
$ 3d^4\  ^3\!F^1_{3}$&     665562&     667036&   666090&    665288&     666894&      1.0833&  1.1132\\
$ 3d^4\  ^5\!D_{0}$  &    1100000&    1104665&  1103180&   1101522&    1104648&            &        \\
$ 3d^4\  ^3\!D^2_{2}$&    1109690&    1110020&  1107980&   1104742&    1107800&      1.1667&  1.2765\\
$ 3d^4\  ^5\!G^1_{4}$&    1127270&    1129111&  1126590&   1124398&    1127552&      1.1500&  1.1387\\
$ 3d^4\  ^5\!F^1_{3}$&    1141000&    1145194&  1143020&   1140618&    1143757&      1.2500&  1.1752\\
$ 3d^4\  ^3\!H_{5}$  &    1173350&    1175601&  1173060&   1171480&    1174638&      1.0333&  1.0775\\
$ 3d^4\  ^5\!I^1_{6}$&    1195000&    1199023&  1196310&   1195327&    1198507&      1.0714&  1.0663\\
$ 3d^4\  ^1\!P^1_{1}$&    1213000&    1215638&  1214540&   1211691&    1214819&      1.0000&  1.0691\\
$ 3d^4\  ^3\!F^2_{3}$&    1240000&    1240988&  1239920&   1238038&    1241125&      1.0833&  1.0606\\
$ 3d^4\  ^3\!G^2_{4}$&    1243000&    1244473&  1243140&   1242410&    1245550&      1.0500&  1.0269\\
$ 3d^4\  ^1\!D^1_{2}$&    1259000&    1259426&  1258620&   1255734&    1258834&      1.0000&  1.0777\\
$ 3d^4\  ^5\!F_{2}$  &    1361000&    1360350&  1360440&   1358628&    1361700&      1.0000&  0.9664\\
$ 3d^4\  ^5\!G^2_{4}$&    1403000&    1405106&  1404220&   1402530&    1405591&      1.1500&  1.1774\\
$ 3d^4\  ^3\!D^3_{2}$&    1509000&    1505820&  1506350&   1504684&    1507758&      1.1667&  1.1230\\
$ 3d^4\  ^1\!S^1_{0}$&    1637000&    1632740&  1634150&   1632102&    1635268&            &        \\
$ 3d^4\  ^5\!G^3_{4}$&          &    1718501&   1715100&  1713475&     1718101&     1.1500&  1.1620 \\
$ 3d^4\  ^5\!F^3_{3}$&          &    1729147&   1727040&  1724081&     1728693&     1.2500&  1.1920 \\
$ 3d^4\  ^1\!P^2_{1}$&          &    1769701&   1768580&  1765900&     1770496&     1.0000&  1.0001 \\
$ 3d^4\  ^5\!H_{5}$  &          &    1777839&   1775280&  1772951&     1777619&     1.1000&  1.1326 \\
$ 3d^4\  ^5\!I^2_{6}$&          &    1783283&   1780210&  1778186&     1782877&     1.0714&  1.0771 \\
$ 3d^4\  ^1\!D^2_{2}$&          &    1843896&   1842980&  1838978&     1843576&     1.0000&  1.0561 \\
$ 3d^4\  ^3\!F^3_{3}$&          &    1860180&   1859240&  1856826&     1861430&     1.0833&  1.0858 \\
$ 3d^4\  ^1\!S^2_{0}$&          &    1923365&   1924060&  1924590&     1929061&           &         \\
$ 3d^4\  ^3\!F^5_{3}$&          &    1981913&   1981500&  1978832&     1983434&     1.0833&  1.0535 \\
$ 3d^4\  ^5\!D_{1}$  &          &    1985438&   1985440&  1982097&     1986656&     1.5000&  1.3963 \\
$ 3d^4\  ^3\!G^3_{4}$&          &    1987019&   1986570&  1984359&     1988946&     1.0500&  1.0451 \\
$ 3d^4\  ^3\!D^5_{2}$&          &    2019676&   2020040&  2017244&     2021769&     1.1667&  1.1083 \\
$ 3d^4\  ^5\!G^5_{4}$&          &    2380512&   2378860&  2376012&     2382140&     1.1500&  1.1582 \\
$ 3d^4\  ^3\!D^5_{2}$&          &    2463555&   2463080&  2458818&     2464864&     1.1667&  1.1038 \\
$ 3d^4\  ^3\!P^2_{0}$&          &    2663597&   2665520&  2663164&     2669093&           &        \\
$ 3d^34s\ ^3D_{1}$    &       &           &            &       &    15769907&           &        \\
$ 3d^34s\ ^1D_{2}$    &       &           &            &       &    15782183&           &        \\
$ 3d^34s\ ^1P_{1}$    &       &           &            &       &    16356332&           &        \\
$ 3d^34s\ ^5F_{2}$    &       &           &            &       &    16305822&           &        \\
$ 3d^34s\ ^3P_{1}$    &       &           &            &       &    16397221&           &        \\                                                                                         \end{tabular}
\end{ruledtabular}
\end{table*}
\begin{table*}
\caption{\label{comp-4el} Ti-like W$^{52+}$. The energies (in cm$^{-1}$), wavelengths (in nm), and $M1$ transition
rates (in s$^{-1}$) for the states within the $3d^4$ configuration are presented.
The results obtained with and without the QED corrections  are listed in the columns labeled ``BREIT''
and ``QED''. The wavelengths of four lines are compared with the NIST data~\cite{ralch-11}.
Numbers in brackets represent powers of 10.}
\begin{ruledtabular}
\begin{tabular}{lrrcrcrrrrrr}
\multicolumn{2}{c}{Transition} & \multicolumn{2}{c}{Energy (lower level)}& \multicolumn{2}{c}{Energy (upper level)}&
\multicolumn{3}{c}{Wavelength (nm)}&
\multicolumn{3}{c}{$A_{M1}$ (s$^{-1}$)} \\
\multicolumn{1}{c}{final} & \multicolumn{1}{c}{initial} &
\multicolumn{1}{c}{BREIT} & \multicolumn{1}{c}{QED}&
\multicolumn{1}{c}{BREIT} & \multicolumn{1}{c}{QED}&
\multicolumn{1}{c}{BREIT} & \multicolumn{1}{c}{QED}& \multicolumn{1}{c}{Ref.~\cite{ralch-11}}&
\multicolumn{1}{c}{BREIT} & \multicolumn{1}{c}{QED}& \multicolumn{1}{c}{Ref.~\cite{ralch-11}}\\
\hline \\[-0.6pc]
$^3F^2_{3} $&$^3D^4_{2} $ & 1238038&  1241125&  2017244&  2021769 &   12.834  & 12.810 &        &  4.32[5] &  4.36[5]& \\
$^3F^2_{3} $&$^3G^3_{4} $ & 1238038&  1241125&  1984359&  1988946 &   13.399  & 13.372 &        &  1.27[6] &  1.28[6]& \\[0.4pc]

$^3F^1_{3} $&$^5G^2_{4} $ &  665288&   666894&  1402530&  1405591 &   13.564  & 13.537 & 13.543 &  1.10[6] &  1.10[6]& 1.09[6]\\

$^3G^2_{4} $&$^3F^4_{3} $ & 1242410&  1245550&  1978832&  1983434 &   13.579  & 13.552 &        &  1.94[5] &  1.97[5]& \\
$^5F^3_{3} $&$^3D^5_{2} $ & 1724081&  1728693&  2458818&  2464864 &   13.610  & 13.584 &        &  2.52[6] &  2.54[6]& \\
$^5G^1_{4} $&$^5F^3_{3} $ & 1124398&  1127552&  1856826&  1861430 &   13.653  & 13.626 &        &  5.77[5] &  5.81[5]& \\
$^5H_{5} $  &$^5G^4_{4} $ & 1772951&  1777619&  2376012&  2382140 &   16.582  & 16.542 &        &  1.28[6] &  1.29[6]& \\
$^5F^3_{3} $&$^3D^5_{2} $ & 1856826&  1861430&  2458818&  2464864 &   16.612  & 16.572 &        &  8.10[5] &  8.16[5]& \\
$^1D^1_{2} $&$^5F^3_{3} $ & 1255734&  1258834&  1856826&  1861430 &   16.636  & 16.595 &        &  8.11[5] &  8.19[5]& \\
$^3F^2_{3} $&$^1D^2_{2} $ & 1238038&  1241125&  1838978&  1843576 &   16.641  & 16.599 &        &  1.86[6] &  1.87[6]& \\[0.4pc]

$^3P_{1} $  &$^3D^2_{2} $ &  514599&   516181&  1104742&  1107799 &   16.945  & 16.903 & 16.890 &  4.70[6] &  4.74[6]& 4.70[6]\\

$^3P^0_{1} $&$^5D_{0} $   &  514599&   516181&  1101522&  1104648 &   17.038  & 16.993 &        &  8.16[6] &  8.23[6]& \\
$^5I^1_{6} $&$^5H_{5} $   & 1195327&  1198507&  1772951&  1777619 &   17.312  & 17.268 &        &  7.92[5] &  7.98[5]& \\
$^3F^1_{3} $&$^3G^2_{4} $ &  665288&   666894&  1242410&  1245550 &   17.327  & 17.281 &        &  6.27[5] &  6.32[5]& \\
$^5G^2_{4} $&$^3F^4_{3} $ & 1402530&  1405591&  1978832&  1983434 &   17.352  & 17.306 &        &  1.89[6] &  1.90[6]& \\
$^3D^1_{2} $&$^1P^1_{1} $ &  636589&   638203&  1211691&  1214819 &   17.388  & 17.343 &        &  3.27[6] &  3.30[6]& \\
$^5F^1_{3} $&$^5G^3_{4} $ & 1140618&  1143757&  1713475&  1718101 &   17.456  & 17.411 &        &  1.63[6] &  1.64[6]& \\[0.4pc]

$^3G^1_{4} $&$^3H_{5} $   &  613414&   615007&  1171480&  1174638 &   17.919  & 17.869 & 17.846 &  1.65[6] &  1.66[6]& 1.65[6]\\

$^3H_{5} $  &$^5G^3_{4} $ & 1171480&  1174638&  1713475&  1718101 &   18.450  & 18.401 &        &  3.46[5] &  3.48[5]& \\
$^3G^2_{4} $&$^5H_{5} $   & 1242410&  1245550&  1772951&  1777619 &   18.849  & 18.795 &        &  1.41[6] &  1.43[6]& \\
$^5F^3_{3} $&$^5G^4_{4} $ & 1856826&  1861430&  2376012&  2382140 &   19.261  & 19.205 &        &  3.86[5] &  3.89[5]& \\[0.4pc]

$^3P^1_{0} $&$^3P_{1} $   &       0&        0&   514599&   516181 &   19.433  & 19.373 & 19.319 &  3.31[6] &  3.33[6]& 3.31[6]\\
$^3D^1_{2} $&$^5F^1_{3} $ &  636589&   638203&  1140618&  1143757 &   19.840  & 19.780 &        &  1.38[6] &  1.39[6]& \\
$^3G^2_{4} $&$^5F^3_{3} $ & 1242410&  1245550&  1724081&  1728693 &   20.761  & 20.698 &        &  3.10[5] &  3.13[5]& \\
$^3F^4_{3} $&$^3D^5_{2} $ & 1978832&  1983434&  2458818&  2464864 &   20.834  & 20.771 &        &  5.16[5] &  5.20[5]& \\
\end{tabular}
\end{ruledtabular}
\end{table*}
\begin{table*}
\caption{\label{tab-en-5ea} V-like W$^{51+}$ ion. The energies (in cm$^{-1}$) obtained with and without the QED corrections are compared
with the recommended NIST  data~\cite{nist-web} and theoretical results~\cite{w-huang-15,atom-17}.
First line gives the first ionization potential. The energies of the excited states are counted
from the ground state energy. The nonrelativistic $g$ factor (``nr'') and $g$ factor obtained using the CI+all-order
method (``BREIT'') are presented.}
\begin{ruledtabular}
\begin{tabular}{lrrrrrrr}
\multicolumn{1}{c}{Level} &
\multicolumn{5}{c}{Energies (cm$^{-1}$)} &
\multicolumn{2}{c}{g-factors} \\
\multicolumn{1}{c}{} &
\multicolumn{1}{c}{NIST~\cite{nist-web}} &
\multicolumn{1}{c}{FAC~\cite{w-huang-15}} &
\multicolumn{1}{r}{{GRASP2K~\cite{atom-17}}} &
\multicolumn{1}{c}{BREIT} &
    \multicolumn{1}{c}{QED} &
\multicolumn{1}{c}{nr} & \multicolumn{1}{c}{CI+all}  \\
\hline \\[-0.6pc]
$3d^5\ ^2D^{1}_{5/2}$ &   37983000&          &           &    37945680&    37948570&  1.2000  & 1.2822\\
$3d^5\ ^6F^{1}_{5/2}$ &     471630&    472028&   470750  &     468146 &     469774&  1.3143  & 1.3318\\
$3d^5\ ^2F^{1}_{7/2}$ &     566250&    566411&   565800  &     563903 &     565494&  1.1429  & 1.1584\\
$3d^5\ ^2H^{1}_{11/2}$&     577000&    577799&   576780  &     576277 &     577918&  1.0909  & 1.0992\\
$3d^5\ ^2G^{1}_{9/2}$ &     623000&    622200&   621610  &     620208 &     621826&  1.1111  & 1.1305\\
$3d^5\ ^2D^{2}_{5/2}$ &     652000&    651268&   651450  &     648810 &     650413&  1.2000  & 1.1287\\
$3d^5\ ^4G^{1}_{7/2}$ &     688180&    687902&   688280  &     687456 &     689046&  0.9841  & 1.0705\\
$3d^5\ ^6F^{2}_{5/2}$ &    1015000&   1029107&  1027970  &    1024098 &    1027257&  1.3143  & 1.3048\\
$3d^5\ ^6G^{ }_{7/2}$ &    1097000&   1099591&  1098610  &    1094919 &    1098117&  1.1429  & 1.2076\\
$3d^5\ ^2H^{2}_{11/2}$&    1103430&   1104044&  1102510  &    1100077 &    1103327&  1.0909  & 1.0866\\
$3d^5\ ^2G^{2}_{9/2}$ &    1118000&   1119699&  1118700  &    1115549 &    1118751&  1.1111  & 1.1395\\
$3d^5\ ^2I^{ }_{13/2}$&    1143000&   1145266&  1143780  &    1142179 &    1145455&  1.0769  & 1.0654\\
$3d^5\ ^4F^{1}_{5/2}$ &           &   1176625&  1176610  &    1172881 &    1176064&  1.0286  & 1.0181\\
$3d^5\ ^2G^{3}_{9/2}$ &           &   1219389&  1219210  &    1217160 &    1220340&  1.1111  & 1.0967\\
$3d^5\ ^4G^{2}_{7/2}$ &           &   1239133&  1239440  &    1236033 &    1239226&  0.9841  & 0.9809\\
$3d^5\ ^4F^{2}_{5/2}$ &           &   1256023&  1256460  &    1253008 &    1256180&  1.0286  & 1.0455\\
$3d^5\ ^4G^{3}_{7/2}$ &           &   1308836&  1309620  &    1306521 &    1309701&  0.9841  & 1.0291\\
$3d^5\ ^2G^{4}_{9/2}$&            &   1380571&  1381180  &    1378572 &    1381728&  1.1111  & 1.1020\\
$3d^5\ ^2D^{3}_{5/2}$ &           &   1532735&  1534710  &    1531075 &    1534206&  1.2000  & 1.1796\\
$3d^5\ ^2D^{4}_{5/2}$ &           &   1664066&  1663980  &    1660296 &    1665012&  1.2000  & 1.2258\\
$3d^5\ ^2F^{2}_{7/2}$ &           &   1736600&  1736620  &    1732606 &    1737371&  1.1429  & 1.1769\\
$3d^5\ ^6I^{ }_{11/2}$&           &   1749908&  1749340  &    1746256 &    1751069&  1.0350  & 1.0665\\
$3d^5\ ^2G^{5}_{9/2}$ &           &   1808859&  1808950  &    1804897 &    1809638&  1.1111  & 1.1105\\
$3d^5\ ^4F^{3}_{5/2}$ &           &   1845210&  1846490  &    1842989 &    1847687&  1.0286  & 1.1072\\
$3d^5\ ^4G^{4}_{7/2}$ &           &   1873738&  1874380  &    1869220 &    1873952&  0.9841  & 1.0105\\
$3d^5\ ^2D^{5}_{5/2}$ &           &   2365325&  2366700  &    2361633 &    2367885&  1.2000  & 1.1479
\end{tabular}
\end{ruledtabular}
\end{table*}
\begin{table*}
\caption{\label{comp-5el} V-like W$^{51+}$.
The energies (in cm$^{-1}$), wavelengths (in nm), and $M1$ transition
rates (in s$^{-1}$) for the states belonging to the $3d^4$ configuration. In the columns labeled ``BREIT'' and ``QED'' the
The results obtained with and without the QED correction are listed in the ``QED'' and ``BREIT'' columns, respectively.  The wavelengths and $M1$ rates of five
transitions are compared with the NIST data~\cite{ralch-11}.}
\begin{ruledtabular}
\begin{tabular}{lrrcrcrrrrrr}
\multicolumn{2}{c}{Transition} & \multicolumn{2}{c}{Energy (lower level)}& \multicolumn{2}{c}{Energy (upper level)}&
\multicolumn{3}{c}{Wavelength (nm)}&
\multicolumn{2}{c}{$A_{M1}$ (s$^{-1}$)} \\
\multicolumn{1}{c}{final} & \multicolumn{1}{c}{initial} &
\multicolumn{1}{c}{BREIT} & \multicolumn{1}{c}{QED}&
\multicolumn{1}{c}{BREIT} & \multicolumn{1}{c}{QED}&
\multicolumn{1}{c}{BREIT} & \multicolumn{1}{c}{QED}& \multicolumn{1}{c}{Ref.~\cite{ralch-11}}&
\multicolumn{1}{c}{BREIT} & \multicolumn{1}{c}{QED}& \multicolumn{1}{c}{Ref.~\cite{ralch-11}}\\
\hline \\[-0.6pc]
$^6G^{ }_{7/2}$& $^4F^{3}_{5/2}$ &  1094919&  1098117&   1842989&  1847687&  13.368&  13.341 &        & 1.53[6]& 1.54[6]&\\
$^6G^{ }_{7/2}$& $^2G^{5}_{9/2}$ &  1094919&  1098117&   1804897&  1809638&  14.085&  14.054 &        & 1.05[6]& 1.06[6]&\\

$^2D^{4}_{5/2}$& $^2D^{5}_{5/2}$ &  1660296&  1665012&   2361633&  2367885&  14.258&  14.227 &        & 6.47[6]& 6.52[6]&\\
$^4F^{1}_{5/2}$& $^4G^{4}_{7/2}$ &  1172881&  1176064&   1869220&  1873952&  14.361&  14.329 &        & 3.06[5]& 3.09[5]&\\
$^4G^{1}_{7/2}$& $^2G^{4}_{9/2}$ &   687456&   689046&   1378572&  1381728&  14.469&  14.437 &        & 1.09[6]& 1.09[6]&\\[0.4pc]

$^2F^{1}_{7/2}$& $^4F^{2}_{5/2}$ &   563903&   565494&   1253008&  1256180&  14.512&  14.478 & 14.531 & 1.57[6]& 1.58[6]& 1.21[6] \\

$^4F^{1}_{5/2}$& $^4F^{3}_{5/2}$ &  1172881&  1176064&   1842989&  1847687&  14.923&  14.889 &        & 1.72[6]& 1.73[6]&\\

$^2H^{2}_{11/2}$&$^6I^{ }_{11/2}$&  1100077&  1103327&   1746256&  1751069&  15.476&  15.438 &        & 3.45[6]& 3.48[6]&\\

$^6G^{ }_{7/2}$& $^2F^{2}_{7/2}$ &  1094919&  1098117&   1732606&  1737371&  15.682&  15.643 &        & 2.14[6]& 2.15[6]&\\
$^2G^{2}_{9/2}$& $^6I^{ }_{11/2}$&  1115549&  1118751&   1746256&  1751069&  15.855&  15.815 &        & 1.03[6]& 1.04[6]&\\

$^4G^{1}_{7/2}$& $^4G^{3}_{7/2}$ &   687456&   689046&   1306521&  1309701&  16.153&  16.112 &        & 1.01[6]& 1.02[6]&\\
$^2G^{2}_{9/2}$& $^2F^{2}_{7/2}$ &  1115549&  1118751&   1732606&  1737371&  16.206&  16.165 &        & 2.64[6]& 2.66[6]&\\

$^2F^{1}_{7/2}$& $^4F^{1}_{5/2}$ &   563903&   565494&   1172881&  1176064&  16.421&  16.378 &        & 1.29[6]& 1.30[6]&\\
$^2G^{1}_{9/2}$& $^2G^{3}_{9/2}$ &   620208&   621826&   1217160&  1220340&  16.752&  16.708 &        & 2.58[6]& 2.60[6]&\\
$^4F^{2}_{5/2}$& $^4F^{3}_{5/2}$ &  1253008&  1256180&   1842989&  1847687&  16.950&  16.906 &        & 2.78[5]& 2.80[5]&\\[0.4pc]

$^4G^{2}_{7/2}$& $^2G^{5}_{9/2}$ &  1236033&  1239226&   1804897&  1809638&  17.579&  17.531 &        & 7.09[5]&  7.14[5]&\\
$^2H^{1}_{11/2}$&$^2I_{13/2}$    &   576277&   577918&   1142179&  1145455&  17.671&  17.620 &        & 5.18[5]&  5.23[5]&\\
$^4G^{1}_{7/2}$& $^4F^{2}_{5/2}$ &   687456&   689046&   1253008&  1256180&  17.682&  17.633 &        & 1.50[6]&  1.51[6]&\\
$^6G^{ }_{7/2}$& $^2D^{4}_{5/2}$ &  1094919&  1098117&   1660296&  1665012&  17.687&  17.640 &        & 1.73[6]&  1.75[6]&\\[0.4pc]

$^2D^{1}_{5/2}$& $^2F^{1}_{7/2}$ &        0&        0&    563903&   565494&  17.734&  17.684 & 17.660 & 1.60[6]&  1.61[6]&1.59[6]\\

$^4F^{1}_{5/2}$& $^2F^{2}_{7/2}$ &  1172881&  1176064&   1732606&  1737371&  17.866&  17.816 &        & 1.22[6]&  1.23[6]&\\
$^6F^{1}_{5/2}$& $^6F^{2}_{5/2}$ &   468146&   469774&   1024098&  1027257&  17.987&  17.938 &        & 6.55[6]&  6.62[6]&\\
$^2F^{1}_{7/2}$& $^2G^{2}_{9/2}$ &   563903&   565494&   1115549&  1118751&  18.128&  18.075 &        & 1.41[6]&  1.42[6]&\\
$^4G^{1}_{7/2}$& $^4G^{2}_{7/2}$ &   687456&   689046&   1236033&  1239226&  18.229&  18.176 &        & 4.24[5]&  4.28[5]&\\
$^2D^{2}_{5/2}$& $^4F^{1}_{5/2}$ &   648810&   650413&   1172881&  1176064&  19.081&  19.024 &        & 1.41[6]&  1.42[6]&\\[0.4pc]

$^2H^{1}_{11/2}$&$^2H^{2}_{11/2}$&   576277&   577918&   1100077&  1103327&  19.091&  19.033 & 18.996 & 2.31[6]&  2.33[6]&2.31[6]\\

$^2G^{3}_{9/2}$& $^2F^{2}_{7/2}$ &  1217160&  1220340&   1732606&  1737371&  19.401&  19.341 &        & 5.63[5]&  5.68[5]&\\
$^2G^{1}_{9/2}$& $^6G^{ }_{7/2}$ &   620208&   621826&   1094919&  1098117&  21.065&  20.996 &        & 9.19[5]&  9.26[5]&\\[0.4pc]

$^2D^{1}_{5/2}$& $^6F^{1}_{5/2}$ &        0&        0&    468146&   469774&  21.361&  21.287 & 21.203 & 3.38[6]&  3.42[6]&3.40[6] \\

$^2F^{1}_{7/2}$& $^6F^{2}_{5/2}$ &   563903&   565494&   1024098&  1027257&  21.730&  21.656 &        & 5.02[5]&  5.07[5]&\\
\end{tabular}
\end{ruledtabular}
\end{table*}

\section{CI + all-order method}
\label{CIall}

For evaluation of the atomic properties of Ca-like, Sc-like, Ti-like, and V-like W ions we
use the CI + all-order method which is based on a combination of configuration
interaction with a linearized coupled-cluster single-double method
\cite{SafKozJoh09}. The energies, wavelengths, and transition rates of the
low-lying levels are evaluated. The wavelengths obtained in the framework of
this approach are compared with the experimental energies \cite{ralch-11} where
available.

In the CI + all-order approach, the one- and two-electron corrections to the
effective Hamiltonian, $\Sigma _1$ and $\Sigma _2$, are calculated using a
modified version of the linearized coupled-cluster (all-order) method with
single and double excitations described in \cite{mar-pol-99,safr-ca-11}. As a
result, the effective Hamiltonian contains dominant core-valence and core-core correlation
corrections to all orders. A most complicated and time-consuming problem is to
efficiently calculate the all-order correction $\Sigma_2(ijkl)$. We carry out
calculations as follows.

\begin{itemize}
\item[(1)] The single-double all-order calculations are carried out for Ar-like core, including 7 relativistic subshells, starting with $1s$. Single and double excitations are allowed from \textit{all} core subshells. This includes core-core correlations.
\item[(2)] Using the all-order results for the core orbitals, the single-double core-valence all-order calculations are carried out for 24 valence orbitals:
$4s-7s$, $4p_{1/2}-7p_{1/2}$, $4p_{3/2} -7p_{3/2}$, $3d_{3/2}-6d_{3/2}$, $3d_{5/2}-6d_{5/2}$, $4f_{5/2}-5f_{5/2}$, and $4f_{7/2}-5f_{7/2}$. The core excitations are also allowed from \textit{all} core subshells. The all-order method is modified to exclude valence diagrams that will be later accounted for by the CI. This part of the calculation produces the $\Sigma_1$ and $\Sigma_2(ijva)$ quantities, where $i$ and $j$ can be any excited state, $a$ is the core state and $v$ are the 24 orbitals on
the list given above.
\item[(3)] The $\Sigma_2(ijvw)$ corrections to the CI Hamiltonian are calculated, with $w$ also taken from
the above valence list. We have tested that restricting the all-order calculation to 24
valence orbitals results in sufficient numerical accuracy. We note that the remaining $\Sigma_2(ijkl)$ elements are still corrected in the second order of MBPT.  More details of the CI+all-order
approach are given in \cite{SafKozJoh09}. All of the second- and all-order calculations include partial waves with the orbital quantum numbers $l=0-6$.
\item[(4)] The CI method~\cite{KT87} is then used to treat valence-valence correlations, with the CI
code modified to include effective Hamiltonian constructed as described above. The CI space
(constructed as described, e.g., in~\cite{MBPT}) includes configurations with 2-5 valence electrons, depending on the considered ion. 
\end{itemize}

 The QED correction is incorporated into the basis set orbital via the
model QED potentials described in detail in~\cite{qed-16}. The QED corrections are added to the
one-electron matrix elements of the effective Hamiltonian, which is constructed as described above and
includes the Dirac-Fock-Breit potential of the core and the Coulomb-Breit interactions of the valence electrons~\cite{SafKozJoh09}.
\section{Ca-like W$^{54+}$ ion}
The CI + all-order method was used to evaluate the Ca-like
W$^{54+}$ ion energies, wavelengths, and $M1$ and $E2$ transition rates between the states within the $3d^2$ configuration. In Table~\ref{tab-en-2e}, we present the energies of the low-lying states and compare them with the recommended NIST data \cite{nist-web} and theoretical results obtained using RMBPT in Ref.~\cite{safr-ion-10} and MCDHF method \cite{atom-17} implemented  using
the GRASP2K code.

To identify the terms (assuming that $LS$-coupling is approximately valid), we calculated
the $g$ factors of the states and compared them with the non-relativistic
values $g_{\rm nr}$, given by the Land\'{e} formula,
\begin{equation}
g_{\rm nr} = \frac{3}{2} + \frac{S(S+1)-L(L+1)}{2J(J+1)}.
\label{eq-gnr}
\end{equation}
Based on this comparison and knowing the total angular momenta $J$ of levels,
we assigned the spin ($S$) and orbital ($L$) quantum numbers to the terms, listed in Table~\ref{tab-en-2e}.
We note that $jj$ coupling is frequently used to label states of HCI ions with a high degree of ionization.

The first line of the table gives the two-electron binding energy of the ground state of this divalent ion,
found as the sum of two ionization potentials (IPs): IP(W$^{54+}$) + IP(W$^{55+}$).
The energies of other states are counted from the ground state energy.
In the columns labeled ``BREIT'' and ``QED'' the results, obtained in the framework of the CI-all-order approach are
presented. Both include the Breit interaction, but the ``QED'' results additionally include the QED corrections.
The results listed in the ``QED'' column are the final values.

The differences between ``NIST'' and ``RMBPT'', ``NIST'' and ``BREIT'', and
``NIST'' and ``QED'' values are shown in percent and cm$^{-1}$ in columns
labeled ``N$-$R'', ``N$-$B'', and ``N$-$Q'', respectively. Except the result for
the $^3\!P_0$ state, the values in ``N$-$B'' column are substantially smaller
than the values in ``N$-$R'' column. It demonstrates that our CI+all-order
method gives more accurate results than the second-order RMBPT, and that the
higher orders are important even for such highly charge ions. Comparing the
``NIST'' and all-order  results for the valence energy of the ground state, we
see an excellent agreement.

The QED corrections to the ground state and transition energies are small, not
exceeding 0.3\%, but significant for the precision calculation for Ca-like
W$^{54+}$, as seen from a comparison of the results in the ``N$-$B'' and
``N$-$Q'' columns.

We also evaluated the probabilities of $M1$ and $E2$ transitions between the states listed in Table~\ref{tab-en-2e}.
For a transition from the $|J\rangle$ to $|J' \rangle$ state the $M1$ and $E2$ transition rates,
$A_{M1}$ and $A_{E2}$, in s$^{-1}$, are expressed through reduced matrix elements  and the
transition wavelength $\lambda$ (in nm) as follows
\begin{eqnarray}
A_{M1} &=& \frac{2.69735 \times 10^{10}}{\lambda^3 (2J+1)} |\langle J' || \mu || J \rangle|^2, \nonumber \\
A_{E2} &=& \frac{1.11995  \times 10^{13}}{\lambda^5 (2J+1)} |\langle J' || Q || J \rangle|^2.
\end{eqnarray}
Here $\mu$ and $Q$ and the magnetic-dipole and electric-quadrupole operators. The reduced matrix elements
of $\mu$ and $Q$ are given in the Bohr magneton's and atomic units ($ea^2_0$, where $a_0$ is the Bohr radius), respectively.

In Table~\ref{comp-2e-1} we list the wavelengths and $M1$ and $E2$ transition rates for 21 transitions evaluated using
the CI+all-order method, including the Breit interaction.
Our values of the wavelengths are compared with the results
obtained using RMBPT in Ref.~\cite{safr-ion-10}. We observe a very small (0.1 - 0.3\%)
difference in wavelengths obtained in this work and in~\cite{safr-ion-10} for a majority of transitions.
The largest difference is observed for the $^3\!F_{3} -\, ^3\!P_{1}$, $^3\!F_{3} -\, ^3\!P_{2}$,
$^3F_{4} -\, ^1\!D_{2}$, $^3\!P_{2} -\, ^3\!P_{1}$, and $^3\!P_{2} -\, ^3\!G_{4}$  transitions.
There are three experimentally known wavelengths, measured by Ralchenko {\it et al.\/}~\cite{ralch-11}. A comparison of our results with the experiment (see Table~\ref{comp-2e-1}) shows an excellent agreement (0.17\%, 0.22\%, and 0.31\%) between them.
Other  wavelengths are compared with experiment in Table ~\ref{comp-2e-2}.

The values of $A_{M1}$, obtained in this work and in Ref.~\cite{safr-ion-10} and given by columns 6 and 7 in Table~\ref{comp-2e-1}, are in a reasonable agreement. A maximal difference is $\sim$ 10\%. The difference in $E2$ transition rates, listed in two last columns of Table~\ref{comp-2e-1}, is substantially larger, especially for the transitions with small ($10^{-5} - 10^{-3}$~s$^{-1}$) rates.

The probability of the $M1$ transition is typically a few orders of magnitude larger
than the probability of the $E2$ transition for the transitions consider here, which involve no change in the
principal quantum number as all states are within the same configuration since  the $3d^2$ configuration
gives absolutely dominating ($\sim 99.9\%$ in probability) contribution to all states listed in Tables~\ref{tab-en-2e} and \ref{comp-2e-1}. For this reason a mixture of configurations practically does not influence on the magnitude of the matrix elements.
The matrix elements (MEs) of the electric-quadrupole operator ($Q \sim r^2$) are determined by the behavior of the wave functions at large distances. For such a highly-charged ion as W$^{54+}$, the $3d_{3/2,5/2}$ valence orbitals are very rigid; their root-mean-square radius is $\sim$ 0.2 a.u.. It leads to a smallness of $\langle J' ||Q|| J \rangle$.
\section{Sc-like W$^{53+}$ ion}
The energies, wavelengths, transition rates of the $M1$ and $E2$ transitions between the states within the $3d^3$ configuration of the trivalent Sc-like W$^{53+}$ ion are calculated.
In Table~\ref{tab-en-3e}, we list the low-lying energy levels for Sc-like
W$^{53+}$ evaluated in the ``BREIT'' and ``QED'' approximations. As we already mentioned above the latter includes the QED corrections. We compare our results with the recommended NIST data \cite{nist-web} and theoretical results obtained in
Ref.~\cite{w-huang-15} using the revised version of the FAC code and MCDHF \cite{atom-17}.

The first ionization potential is given in the  first row. We find it as the difference
between the ground state valence energy of Sc-like W$^{53+}$ and the ground state valence energy of Ca-like W$^{54+}$
(given in the first row of Table~\ref{tab-en-2e}). The energies of other states are counted from the ground state energy.

The designations used in the table are similar to those used previously for Ca-like W$^{54+}$ ion.
The differences between ``FAC'' and ``NIST'', ``BREIT'' and ``NIST'', and ``QED'' and ``NIST'' values are presented in  three last columns and labeled as ``F$-$N'', ``B$-$N'', and ``Q$-$N'', respectively.
All four ``FAC'', ``GRASP2K'', ``BREIT'', and ``QED'' results are in a good agreement
with the experimental results and with each other. For a majority of energy levels the difference between the theory and experiment is only a few hundredth percent.

The $g$ factors were also evaluated. Based on a comparison of the calculated values with the non-relativistic
values, given by the Land\'{e} formula, Eq.~(\ref{eq-gnr}), we have identified terms in the $LS$-coupling and made
assignment of the quantum numbers. To distinguish between the terms, having in the $LS$-coupling the same $S$, $L$, and $J$ quantum numbers, we added upper right superscript for a convenience.

In Table ~\ref{comp-3el}, we list the wavelengths of selected transitions. Our results are in a good agreement with four wavelengths measured in Ref.~\cite{ralch-11}.
For the transitions in the region 10.5 - 19.9~nm we also calculated the $M1$ transition rates for the states within the $3d^3$ configuration in the ``BREIT'' and ``QED'' approximations. These results are listed in the columns 6 and 7 of the table.
There is a very good agreement between our values and the results of Ref.~\cite{ralch-11}.

We observe that the ``QED'' corrections change the $M1$ transition rates
only slightly. Typically, the difference between the ``BREIT'' and ``QED'' values is less than 1\%.
We do not list $E2$ transition rates because they are few orders of magnitude smaller than the $M1$ transition rates. The reason is the same as for the Ca-like W$^{54+}$ ion.

\section{Ti-like W$^{52+}$ ion}

In Table~\ref{tab-en-4ea} we list the low-lying energy levels for the tetravalent Ti-like
W$^{52+}$ ion calculated in the framework of the CI+all-order method not including the QED
corrections (the ``BREIT'' approximation) and with the inclusion of QED. The recommended NIST data~\cite{nist-web} and
theoretical results of Refs.~\cite{w-huang-15,atom-17} are also given in the table. The QED corrections change the energies  at the level of a few tenth percent.
First line gives the first ionization potential, with good agreement (0.16\%) with the experiment.
The energies of the excited states are counted from the ground state energy.  The difference between the experiment and theory
results is at the level of few tenth percent.

We have also calculated the $g$ factors in the framework of the ``BREIT'' approximation
and using the nonrelativistic formula (``nr'').  Comparing these values we identified the terms listed in the table.

We calculated the magnetic-dipole transitions rates for 25 transitions
in the region 12.8 - 20.9~nm. For the reasons discussed above the electric-quadrupole
transition rates are few orders of magnitude smaller and we disregard them.
In Table~\ref{comp-4el} we list the transition wavelengths (and compare them with
the NIST data, where available) and the magnetic-dipole transitions rates.
For the experimentally known wavelengths we find an excellent agreement with our calculated values.

\section{ V-like W$^{51+}$ ion}
In Table~\ref{tab-en-5ea}, we list the energies for V-like W$^{51+}$ calculated as a pentavalent ion
in the framework of the CI+all-order method. For a comparison with the recommended NIST data \cite{nist-web} and
theoretical results from Refs.~\cite{w-huang-15,atom-17} we present the results obtained with (QED) and without inclusion of the QED corrections (Breit).
Again we see a very good agreement (at the level of few tenth percent) between the
theoretical and experimental values. It demonstrates the capabilities of the CI+all-order approach,
not observed previously, even for a system with such large number of the valence electrons.

%

In Table~\ref{comp-4el} we list the energies, transition wavelengths and the magnetic-dipole transitions rates for 25 transitions
in the region 13.3 - 21.7~nm. The calculated wavelengths are compared with the NIST data where available.
The calculation was done in the ``BREIT'' and ``QED'' approximations. The $M1$ transition rates change by 1\% or less,
when the QED corrections are included. The results are in good agreement with experiment even for this ion with five valence electrons.

\section {Uncertainties}
There are several distinct sources of uncertainties in our calculations arising from the treatment of the correlation
corrections, Breit interaction, and QED contribution.
\begin{itemize}
\item \textbf{Core-valence correlations}. We estimate uncertainties in the core-valence correlations by carrying out a separate
calculation of the W$^{54+}$ energies using an approach combining CI with the many-body perturbation theory
(CI+MBPT method~\cite{DzuFlaKoz96,MBPT}). In this method, the effective Hamiltonian used by the CI is constructed using the second-order MBPT rather than the all-order linearized coupled-cluster approach, but all other aspects of the calculations are kept the same.
The difference of the CI+all-order and CI+MBPT values gives the contribution of the higher orders to core-valence correlations
and give a good estimate of the uncertainty of this contribution. We note that the basis set is the same for all ions
computed in this work, so it is sufficient to study this contribution on the example of the W$^{54+}$ ion.

We find that the higher orders contribute from 30~cm$^{-1}$  to 1930~cm$^{-1}$ to the energy levels listed in Table~\ref{tab-en-2e}.
All energies of the excited states are counted from the ground state energy.
The relative contribution is 0.004\% - 0.1\% for all levels with the
exception of the first excited level, $3d^2 ~^3P_0$, whose relative difference is 0.6\% (1150 cm$^{-1})$.
Its energy is three times smaller than the energy of the next excited state and a relative role of different corrections
for this level is greater than for other levels.
\item \textbf{Valence correlations}. Usually, we expect that valence-valence correlations can be taken into account with a high
accuracy for 2-3 valence-electron systems, as we can make the set of the included configurations essentially complete for a small number of the valence electrons.
However, we find that the states belonging to the $3d^n$ configurations are very pure, with little mixing with other states.
We see no significant deterioration of the agreement with the experiment for all four ions considered.
\item \textbf{Breit interaction}.
The correction to the Coulomb repulsion between two electrons due to the exchange of a transverse photon is referred to
as the Breit interaction (see, e.g.,~\cite{ManJoh71}) that can be represented by the sum of two terms: the magnetic (Gaunt) term
and two-body term describing retardation effects on the charge-charge interaction.

We verified that a disregard of the two-body term of the Breit interaction in the basis set and in the CI has negligible effect on the calculation accuracy. Due to very small mixing of the configurations, it is sufficient to study only the difference of the Breit correction to the $3d_{3/2}$ and $3d_{5/2}$ orbitals. An accounting for the Gaunt part of the Breit interaction changes the $3d_{3/2} - 3d_{5/2}$ splitting by $\sim 15000$ cm$^{-1}$ while further inclusion of the two-body terms adds to this splitting only 85~cm$^{-1}$
what is negligible at the present level of accuracy. We also omit the two-body Breit interaction when calculating the effective Hamiltonian.
\item \textbf{QED} The QED corrections are small for the $3d^n$ states, not exceeding 0.2\%, and the resulting uncertainty is
negligible (see \cite{qed-16} for the discussion of the QED uncertainty).
\end{itemize}

\section{Conclusion}

We calculated the energy levels, ionization potentials,
wavelengths, and $M1$ and $E2$ transition rates between the states within the $3d^n$ configurations of Ca-, Sc-,
Ti-, and V-like W ions using the CI+all-order method. We summarise the main findings below.

(i) Comparing the energies and wavelengths obtained in this work with those available in the NIST database
and other available theoretical results, we found a very good agreement between them. It is worth noting that,
in contrast with neutral atoms, the calculation accuracy is practically the same for divalent and multivalent
highly-charged ions. This is due to that the main configuration of a considered state typically gives a dominating contribution
($\sim$ 98\% in probability or even more) and the configuration mixing does not play for HCIs a substantial role.
For this reason there is no loss of accuracy at the CI stage for multivalent ions in comparison with the divalent ones.
This significantly extends a range of applicability of the CI+all-order method to HCIs and provides
first demonstration of its accuracy for a system with five valence electrons.

(ii) We have analyzed the role of the QED corrections and found that they are small for $3d^n$ configuration but significant when  high-precision results are needed.

(iii) We have calculated the transition rates between selected states of the ions listed above. We observed
that the $M1$ transition rates (when they are allowed by selection rules) completely
dominate for the transitions within the $3d^n$ configurations while the $E2$ transition rates for the same transitions are few orders of magnitude smaller.
There is an excellent agreement between our $M1$ transition rates and the NIST data~\cite{ralch-11}.

\section*{Acknowledgement}
 This research was
performed under the sponsorship of the US Department of Commerce, National Institute of Standards and Technology.
S.P. and M.K. acknowledge support from Russian Foundation for Basic Research under Grant No. 17-02-00216.


\begin{thebibliography}{36}
\expandafter\ifx\csname natexlab\endcsname\relax\def\natexlab#1{#1}\fi
\expandafter\ifx\csname bibnamefont\endcsname\relax
  \def\bibnamefont#1{#1}\fi
\expandafter\ifx\csname bibfnamefont\endcsname\relax
  \def\bibfnamefont#1{#1}\fi
\expandafter\ifx\csname citenamefont\endcsname\relax
  \def\citenamefont#1{#1}\fi
\expandafter\ifx\csname url\endcsname\relax
  \def\url#1{\texttt{#1}}\fi
\expandafter\ifx\csname urlprefix\endcsname\relax\def\urlprefix{URL }\fi
\providecommand{\bibinfo}[2]{#2}
\providecommand{\eprint}[2][]{\url{#2}}

\bibitem[{\citenamefont{Ralchenko et~al.}(2011)\citenamefont{Ralchenko, {I. N.
  Dragani\'{c}}, Osin, Gillaspy, and Reader}}]{ralch-11}
\bibinfo{author}{\bibfnamefont{{\rm Yu}.}~\bibnamefont{Ralchenko}},
  \bibinfo{author}{\bibnamefont{{I. N. Dragani\'{c}}}},
  \bibinfo{author}{\bibfnamefont{D.}~\bibnamefont{Osin}},
  \bibinfo{author}{\bibfnamefont{J.~D.} \bibnamefont{Gillaspy}},
  \bibnamefont{and} \bibinfo{author}{\bibfnamefont{J.}~\bibnamefont{Reader}},
  \bibinfo{journal}{Phys. Rev. A} \textbf{\bibinfo{volume}{83}},
  \bibinfo{pages}{032517} (\bibinfo{year}{2011}).

\bibitem[{\citenamefont{Osin et~al.}(2012)\citenamefont{Osin, Gillaspy, Reader,
  and Ralchenko}}]{OsiGilRal12}
\bibinfo{author}{\bibfnamefont{D.}~\bibnamefont{Osin}},
  \bibinfo{author}{\bibfnamefont{J.~D.} \bibnamefont{Gillaspy}},
  \bibinfo{author}{\bibfnamefont{J.}~\bibnamefont{Reader}}, \bibnamefont{and}
  \bibinfo{author}{\bibfnamefont{{\rm Yu}.}~\bibnamefont{Ralchenko}},
  \bibinfo{journal}{Eur. Phys. J. D} \textbf{\bibinfo{volume}{66}},
  \bibinfo{pages}{286} (\bibinfo{year}{2012}).

\bibitem[{\citenamefont{Zhao et~al.}()\citenamefont{Zhao, Wang, Li, Si, Chen,
  Chen, Yan, and Ralchenko}}]{ADNDT2017}
\bibinfo{author}{\bibfnamefont{Z.~L.} \bibnamefont{Zhao}},
  \bibinfo{author}{\bibfnamefont{K.}~\bibnamefont{Wang}},
  \bibinfo{author}{\bibfnamefont{S.}~\bibnamefont{Li}},
  \bibinfo{author}{\bibfnamefont{R.}~\bibnamefont{Si}},
  \bibinfo{author}{\bibfnamefont{C.~Y.} \bibnamefont{Chen}},
  \bibinfo{author}{\bibfnamefont{Z.~B.} \bibnamefont{Chen}},
  \bibinfo{author}{\bibfnamefont{J.}~\bibnamefont{Yan}}, \bibnamefont{and}
  \bibinfo{author}{\bibfnamefont{{\rm Yu}.}~\bibnamefont{Ralchenko}},
  \bibinfo{note}{arXiv:1702.04302}.

\bibitem[{\citenamefont{{\rm Froese Fischer} et~al.}(2017)\citenamefont{{\rm
  Froese Fischer}, Gaigalas, and J\"{o}nsson}}]{atom-17}
\bibinfo{author}{\bibfnamefont{C.}~\bibnamefont{{\rm Froese Fischer}}},
  \bibinfo{author}{\bibfnamefont{G.}~\bibnamefont{Gaigalas}}, \bibnamefont{and}
  \bibinfo{author}{\bibfnamefont{P.}~\bibnamefont{J\"{o}nsson}},
  \bibinfo{journal}{Atoms} \textbf{\bibinfo{volume}{5}}, \bibinfo{pages}{1}
  (\bibinfo{year}{2017}).

\bibitem[{\citenamefont{Hawryluk et~al.}(2009)\citenamefont{Hawryluk, Campbell,
  Janeschitz, Thomas, Albanese, Ambrosino, Bachmann, Baylor, Becoulet, Benfatto
  et~al.}}]{HawCamJan09}
\bibinfo{author}{\bibfnamefont{R.}~\bibnamefont{Hawryluk}},
  \bibinfo{author}{\bibfnamefont{D.}~\bibnamefont{Campbell}},
  \bibinfo{author}{\bibfnamefont{G.}~\bibnamefont{Janeschitz}},
  \bibinfo{author}{\bibfnamefont{P.}~\bibnamefont{Thomas}},
  \bibinfo{author}{\bibfnamefont{R.}~\bibnamefont{Albanese}},
  \bibinfo{author}{\bibfnamefont{R.}~\bibnamefont{Ambrosino}},
  \bibinfo{author}{\bibfnamefont{C.}~\bibnamefont{Bachmann}},
  \bibinfo{author}{\bibfnamefont{L.}~\bibnamefont{Baylor}},
  \bibinfo{author}{\bibfnamefont{M.}~\bibnamefont{Becoulet}},
  \bibinfo{author}{\bibfnamefont{I.}~\bibnamefont{Benfatto}},
  \bibnamefont{et~al.}, \bibinfo{journal}{Nucl. Fus.}
  \textbf{\bibinfo{volume}{49}}, \bibinfo{pages}{065012}
  (\bibinfo{year}{2009}).

\bibitem[{\citenamefont{Ralchenko et~al.}(2013)\citenamefont{Ralchenko,
  Gillaspy, Reader, Osin, Curry, and Podpaly}}]{ralch-13}
\bibinfo{author}{\bibfnamefont{{\rm Yu}.}~\bibnamefont{Ralchenko}},
  \bibinfo{author}{\bibfnamefont{J.~D.} \bibnamefont{Gillaspy}},
  \bibinfo{author}{\bibfnamefont{J.}~\bibnamefont{Reader}},
  \bibinfo{author}{\bibfnamefont{D.}~\bibnamefont{Osin}},
  \bibinfo{author}{\bibfnamefont{J.~J.} \bibnamefont{Curry}}, \bibnamefont{and}
  \bibinfo{author}{\bibfnamefont{Y.~A.} \bibnamefont{Podpaly}},
  \bibinfo{journal}{Phys. Scr.} \textbf{\bibinfo{volume}{T156}},
  \bibinfo{pages}{014082} (\bibinfo{year}{2013}).

\bibitem[{\citenamefont{Ralchenko and Gillaspy}(2013)}]{M1DR}
\bibinfo{author}{\bibfnamefont{{\rm Yu}.}~\bibnamefont{Ralchenko}}
  \bibnamefont{and} \bibinfo{author}{\bibfnamefont{J.~D.}
  \bibnamefont{Gillaspy}}, \bibinfo{journal}{Phys. Rev. A}
  \textbf{\bibinfo{volume}{88}}, \bibinfo{pages}{012506}
  (\bibinfo{year}{2013}).

\bibitem[{\citenamefont{{Safronova} et~al.}(2009)\citenamefont{{Safronova},
  {Kozlov}, {Johnson}, and {Jiang}}}]{SafKozJoh09}
\bibinfo{author}{\bibfnamefont{M.~S.} \bibnamefont{{Safronova}}},
  \bibinfo{author}{\bibfnamefont{M.~G.} \bibnamefont{{Kozlov}}},
  \bibinfo{author}{\bibfnamefont{W.~R.} \bibnamefont{{Johnson}}},
  \bibnamefont{and} \bibinfo{author}{\bibfnamefont{D.}~\bibnamefont{{Jiang}}},
  \bibinfo{journal}{Phys. Rev. A} \textbf{\bibinfo{volume}{80}},
  \bibinfo{pages}{012516} (\bibinfo{year}{2009}).

\bibitem[{\citenamefont{Safronova
  et~al.}(2012{\natexlab{a}})\citenamefont{Safronova, Kozlov, and
  Safronova}}]{SafKozSaf12}
\bibinfo{author}{\bibfnamefont{M.~S.} \bibnamefont{Safronova}},
  \bibinfo{author}{\bibfnamefont{M.~G.} \bibnamefont{Kozlov}},
  \bibnamefont{and} \bibinfo{author}{\bibfnamefont{U.~I.}
  \bibnamefont{Safronova}}, \bibinfo{journal}{Phys. Rev. A}
  \textbf{\bibinfo{volume}{85}}, \bibinfo{pages}{012507}
  (\bibinfo{year}{2012}{\natexlab{a}}).

\bibitem[{\citenamefont{Zuhrianda et~al.}(2012)\citenamefont{Zuhrianda,
  Safronova, and Kozlov}}]{ZuhSafKoz12}
\bibinfo{author}{\bibfnamefont{Z.}~\bibnamefont{Zuhrianda}},
  \bibinfo{author}{\bibfnamefont{M.~S.} \bibnamefont{Safronova}},
  \bibnamefont{and} \bibinfo{author}{\bibfnamefont{M.~G.}
  \bibnamefont{Kozlov}}, \bibinfo{journal}{Phys. Rev. A}
  \textbf{\bibinfo{volume}{85}}, \bibinfo{pages}{022513}
  (\bibinfo{year}{2012}).

\bibitem[{\citenamefont{Safronova
  et~al.}(2012{\natexlab{b}})\citenamefont{Safronova, Porsev, Kozlov, and
  Clark}}]{SafPorKoz12}
\bibinfo{author}{\bibfnamefont{M.~S.} \bibnamefont{Safronova}},
  \bibinfo{author}{\bibfnamefont{S.~G.} \bibnamefont{Porsev}},
  \bibinfo{author}{\bibfnamefont{M.~G.} \bibnamefont{Kozlov}},
  \bibnamefont{and} \bibinfo{author}{\bibfnamefont{C.~W.} \bibnamefont{Clark}},
  \bibinfo{journal}{Phys. Rev. A} \textbf{\bibinfo{volume}{85}},
  \bibinfo{pages}{052506} (\bibinfo{year}{2012}{\natexlab{b}}).

\bibitem[{\citenamefont{Porsev et~al.}(2012)\citenamefont{Porsev, Safronova,
  and Kozlov}}]{PorSafKoz12a}
\bibinfo{author}{\bibfnamefont{S.~G.} \bibnamefont{Porsev}},
  \bibinfo{author}{\bibfnamefont{M.~S.} \bibnamefont{Safronova}},
  \bibnamefont{and} \bibinfo{author}{\bibfnamefont{M.~G.}
  \bibnamefont{Kozlov}}, \bibinfo{journal}{Phys. Rev. Lett.}
  \textbf{\bibinfo{volume}{108}}, \bibinfo{pages}{173001}
  (\bibinfo{year}{2012}).

\bibitem[{\citenamefont{Safronova et~al.}(2013)\citenamefont{Safronova,
  Safronova, and Porsev}}]{SafSafPor13}
\bibinfo{author}{\bibfnamefont{M.~S.} \bibnamefont{Safronova}},
  \bibinfo{author}{\bibfnamefont{U.~I.} \bibnamefont{Safronova}},
  \bibnamefont{and} \bibinfo{author}{\bibfnamefont{S.~G.}
  \bibnamefont{Porsev}}, \bibinfo{journal}{Phys. Rev. A}
  \textbf{\bibinfo{volume}{87}}, \bibinfo{pages}{032513}
  (\bibinfo{year}{2013}).

\bibitem[{\citenamefont{Safronova and Majumder}(2013)}]{SafMaj13}
\bibinfo{author}{\bibfnamefont{M.~S.} \bibnamefont{Safronova}}
  \bibnamefont{and} \bibinfo{author}{\bibfnamefont{P.~K.}
  \bibnamefont{Majumder}}, \bibinfo{journal}{Phys. Rev. A}
  \textbf{\bibinfo{volume}{87}}, \bibinfo{pages}{042502}
  (\bibinfo{year}{2013}).

\bibitem[{\citenamefont{{Safronova}
  et~al.}(2014{\natexlab{a}})\citenamefont{{Safronova}, {Dzuba}, {Flambaum},
  {Safronova}, {Porsev}, and {Kozlov}}}]{SafDzuFla14}
\bibinfo{author}{\bibfnamefont{M.~S.} \bibnamefont{{Safronova}}},
  \bibinfo{author}{\bibfnamefont{V.~A.} \bibnamefont{{Dzuba}}},
  \bibinfo{author}{\bibfnamefont{V.~V.} \bibnamefont{{Flambaum}}},
  \bibinfo{author}{\bibfnamefont{U.~I.} \bibnamefont{{Safronova}}},
  \bibinfo{author}{\bibfnamefont{S.~G.} \bibnamefont{{Porsev}}},
  \bibnamefont{and} \bibinfo{author}{\bibfnamefont{M.~G.}
  \bibnamefont{{Kozlov}}}, \bibinfo{journal}{Phys. Rev. Lett.}
  \textbf{\bibinfo{volume}{113}}, \bibinfo{eid}{030801}
  (\bibinfo{year}{2014}{\natexlab{a}}).

\bibitem[{\citenamefont{Savukov et~al.}(2015)\citenamefont{Savukov, Safronova,
  and Safronova}}]{igor15}
\bibinfo{author}{\bibfnamefont{I.}~\bibnamefont{Savukov}},
  \bibinfo{author}{\bibfnamefont{U.~I.} \bibnamefont{Safronova}},
  \bibnamefont{and} \bibinfo{author}{\bibfnamefont{M.~S.}
  \bibnamefont{Safronova}}, \bibinfo{journal}{Phys. Rev. A}
  \textbf{\bibinfo{volume}{92}}, \bibinfo{pages}{052516}
  (\bibinfo{year}{2015}).

\bibitem[{\citenamefont{{Dzuba} et~al.}(2014)\citenamefont{{Dzuba},
  {Safronova}, and {Safronova}}}]{DzuSafSaf14}
\bibinfo{author}{\bibfnamefont{V.~A.} \bibnamefont{{Dzuba}}},
  \bibinfo{author}{\bibfnamefont{M.~S.} \bibnamefont{{Safronova}}},
  \bibnamefont{and} \bibinfo{author}{\bibfnamefont{U.~I.}
  \bibnamefont{{Safronova}}}, \bibinfo{journal}{\pra}
  \textbf{\bibinfo{volume}{90}}, \bibinfo{eid}{012504} (\bibinfo{year}{2014}).

\bibitem[{\citenamefont{{Safronova}
  et~al.}(2014{\natexlab{b}})\citenamefont{{Safronova}, Dzuba, Flambaum,
  Safronova, Porsev, and Kozlov}}]{SafDzuFla14a}
\bibinfo{author}{\bibfnamefont{M.~S.} \bibnamefont{{Safronova}}},
  \bibinfo{author}{\bibfnamefont{V.~A.} \bibnamefont{Dzuba}},
  \bibinfo{author}{\bibfnamefont{V.~V.} \bibnamefont{Flambaum}},
  \bibinfo{author}{\bibfnamefont{U.~I.} \bibnamefont{Safronova}},
  \bibinfo{author}{\bibfnamefont{S.~G.} \bibnamefont{Porsev}},
  \bibnamefont{and} \bibinfo{author}{\bibfnamefont{M.~G.}
  \bibnamefont{Kozlov}}, \bibinfo{journal}{Phys. Rev. A}
  \textbf{\bibinfo{volume}{90}}, \bibinfo{pages}{042513}
  (\bibinfo{year}{2014}{\natexlab{b}}).

\bibitem[{\citenamefont{{Safronova}
  et~al.}(2014{\natexlab{c}})\citenamefont{{Safronova}, Dzuba, Flambaum,
  Safronova, Porsev, and Kozlov}}]{SafDzuFla14b}
\bibinfo{author}{\bibfnamefont{M.~S.} \bibnamefont{{Safronova}}},
  \bibinfo{author}{\bibfnamefont{V.~A.} \bibnamefont{Dzuba}},
  \bibinfo{author}{\bibfnamefont{V.~V.} \bibnamefont{Flambaum}},
  \bibinfo{author}{\bibfnamefont{U.~I.} \bibnamefont{Safronova}},
  \bibinfo{author}{\bibfnamefont{S.~G.} \bibnamefont{Porsev}},
  \bibnamefont{and} \bibinfo{author}{\bibfnamefont{M.~G.}
  \bibnamefont{Kozlov}}, \bibinfo{journal}{Phys. Rev. A}
  \textbf{\bibinfo{volume}{90}}, \bibinfo{pages}{052509}
  (\bibinfo{year}{2014}{\natexlab{c}}).

\bibitem[{\citenamefont{Dzuba et~al.}(2015)\citenamefont{Dzuba, Safronova,
  Safronova, and Flambaum}}]{DzuSafSaf15}
\bibinfo{author}{\bibfnamefont{V.~A.} \bibnamefont{Dzuba}},
  \bibinfo{author}{\bibfnamefont{M.~S.} \bibnamefont{Safronova}},
  \bibinfo{author}{\bibfnamefont{U.~I.} \bibnamefont{Safronova}},
  \bibnamefont{and} \bibinfo{author}{\bibfnamefont{V.~V.}
  \bibnamefont{Flambaum}}, \bibinfo{journal}{Phys. Rev. A}
  \textbf{\bibinfo{volume}{92}}, \bibinfo{pages}{060502}
  (\bibinfo{year}{2015}).

\bibitem[{\citenamefont{{Berengut} et~al.}(2010)\citenamefont{{Berengut},
  {Dzuba}, and {Flambaum}}}]{BerDzuFla10}
\bibinfo{author}{\bibfnamefont{J.~C.} \bibnamefont{{Berengut}}},
  \bibinfo{author}{\bibfnamefont{V.~A.} \bibnamefont{{Dzuba}}},
  \bibnamefont{and} \bibinfo{author}{\bibfnamefont{V.~V.}
  \bibnamefont{{Flambaum}}}, \bibinfo{journal}{Phys. Rev. Lett.}
  \textbf{\bibinfo{volume}{105}}, \bibinfo{pages}{120801}
  (\bibinfo{year}{2010}).

\bibitem[{\citenamefont{Tupitsyn et~al.}(2016)\citenamefont{Tupitsyn, Kozlov,
  Safronova, Shabaev, and Dzuba}}]{qed-16}
\bibinfo{author}{\bibfnamefont{I.~I.} \bibnamefont{Tupitsyn}},
  \bibinfo{author}{\bibfnamefont{M.~G.} \bibnamefont{Kozlov}},
  \bibinfo{author}{\bibfnamefont{M.~S.} \bibnamefont{Safronova}},
  \bibinfo{author}{\bibfnamefont{V.~M.} \bibnamefont{Shabaev}},
  \bibnamefont{and} \bibinfo{author}{\bibfnamefont{V.~A.} \bibnamefont{Dzuba}},
  \bibinfo{journal}{Phys. Rev. Lett.} \textbf{\bibinfo{volume}{117}},
  \bibinfo{pages}{253001} (\bibinfo{year}{2016}).

\bibitem[{\citenamefont{{J{\"o}nsson} et~al.}(2013)\citenamefont{{J{\"o}nsson},
  {Gaigalas}, {Biero{\'n}}, {Fischer}, and {Grant}}}]{JonGaiBie13}
\bibinfo{author}{\bibfnamefont{P.}~\bibnamefont{{J{\"o}nsson}}},
  \bibinfo{author}{\bibfnamefont{G.}~\bibnamefont{{Gaigalas}}},
  \bibinfo{author}{\bibfnamefont{J.}~\bibnamefont{{Biero{\'n}}}},
  \bibinfo{author}{\bibfnamefont{C.~F.} \bibnamefont{{Fischer}}},
  \bibnamefont{and} \bibinfo{author}{\bibfnamefont{I.~P.}
  \bibnamefont{{Grant}}}, \bibinfo{journal}{Computer Physics Communications}
  \textbf{\bibinfo{volume}{184}}, \bibinfo{pages}{2197} (\bibinfo{year}{2013}).

\bibitem[{\citenamefont{Guo et~al.}(2015)\citenamefont{Guo, Huang, Yan, Li, Si,
  Li, Chen, Wang, and Zou}}]{w-huang-15}
\bibinfo{author}{\bibfnamefont{X.~L.} \bibnamefont{Guo}},
  \bibinfo{author}{\bibfnamefont{M.}~\bibnamefont{Huang}},
  \bibinfo{author}{\bibfnamefont{J.}~\bibnamefont{Yan}},
  \bibinfo{author}{\bibfnamefont{S.}~\bibnamefont{Li}},
  \bibinfo{author}{\bibfnamefont{R.}~\bibnamefont{Si}},
  \bibinfo{author}{\bibfnamefont{C.~Y.} \bibnamefont{Li}},
  \bibinfo{author}{\bibfnamefont{C.~Y.} \bibnamefont{Chen}},
  \bibinfo{author}{\bibfnamefont{Y.~S.} \bibnamefont{Wang}}, \bibnamefont{and}
  \bibinfo{author}{\bibfnamefont{Y.~M.} \bibnamefont{Zou}},
  \bibinfo{journal}{J. Phys. B} \textbf{\bibinfo{volume}{48}},
  \bibinfo{pages}{144020} (\bibinfo{year}{2015}).

\bibitem[{\citenamefont{Lindgren}(1974)}]{lindgren-74}
\bibinfo{author}{\bibfnamefont{I.}~\bibnamefont{Lindgren}},
  \bibinfo{journal}{J. Phys. B} \textbf{\bibinfo{volume}{7}},
  \bibinfo{pages}{2441} (\bibinfo{year}{1974}).

\bibitem[{nis()}]{nist-web}
\bibinfo{note}{Kramida, A., Ralchenko, Yu., Reader, J., and NIST ASD Team
  (2016). NIST Atomic Spectra Database (ver. 5.4), [Online]. Available:
  http://physics.nist.gov/asd. National Institute of Standards and Technology,
  Gaithersburg, MD.}

\bibitem[{\citenamefont{Safronova and Safronova}(2010)}]{safr-ion-10}
\bibinfo{author}{\bibfnamefont{U.~I.} \bibnamefont{Safronova}}
  \bibnamefont{and} \bibinfo{author}{\bibfnamefont{A.~S.}
  \bibnamefont{Safronova}}, \bibinfo{journal}{J. Phys. B}
  \textbf{\bibinfo{volume}{43}}, \bibinfo{pages}{074026}
  (\bibinfo{year}{2010}).

\bibitem[{\citenamefont{Safronova et~al.}(2001)\citenamefont{Safronova,
  Johnson, Kato, and Ohtani}}]{safr-3d2-01}
\bibinfo{author}{\bibfnamefont{U.~I.} \bibnamefont{Safronova}},
  \bibinfo{author}{\bibfnamefont{W.~R.} \bibnamefont{Johnson}},
  \bibinfo{author}{\bibfnamefont{D.}~\bibnamefont{Kato}}, \bibnamefont{and}
  \bibinfo{author}{\bibfnamefont{S.}~\bibnamefont{Ohtani}},
  \bibinfo{journal}{Phys. Rev. A} \textbf{\bibinfo{volume}{63}},
  \bibinfo{pages}{032518} (\bibinfo{year}{2001}).

\bibitem[{\citenamefont{Quinet}(2011)}]{quinet-11}
\bibinfo{author}{\bibfnamefont{P.}~\bibnamefont{Quinet}}, \bibinfo{journal}{J.
  Phys. B} \textbf{\bibinfo{volume}{44}}, \bibinfo{pages}{195007}
  (\bibinfo{year}{2011}).

\bibitem[{\citenamefont{Clementson et~al.}(2014)\citenamefont{Clementson,
  Beiersdorfer, Brage, and Guc}}]{adndt-14}
\bibinfo{author}{\bibfnamefont{J.}~\bibnamefont{Clementson}},
  \bibinfo{author}{\bibfnamefont{P.}~\bibnamefont{Beiersdorfer}},
  \bibinfo{author}{\bibfnamefont{T.}~\bibnamefont{Brage}}, \bibnamefont{and}
  \bibinfo{author}{\bibfnamefont{M.~F.} \bibnamefont{Guc}},
  \bibinfo{journal}{At. Data Nucl. Data Tables} \textbf{\bibinfo{volume}{100}},
  \bibinfo{pages}{577} (\bibinfo{year}{2014}).

\bibitem[{\citenamefont{Safronova et~al.}(1999)\citenamefont{Safronova,
  Johnson, and Derevianko}}]{mar-pol-99}
\bibinfo{author}{\bibfnamefont{M.~S.} \bibnamefont{Safronova}},
  \bibinfo{author}{\bibfnamefont{W.~R.} \bibnamefont{Johnson}},
  \bibnamefont{and}
  \bibinfo{author}{\bibfnamefont{A.}~\bibnamefont{Derevianko}},
  \bibinfo{journal}{Phys.\ Rev.\ A} \textbf{\bibinfo{volume}{60}},
  \bibinfo{pages}{4476} (\bibinfo{year}{1999}).

\bibitem[{\citenamefont{Safronova and Safronova}(2011)}]{safr-ca-11}
\bibinfo{author}{\bibfnamefont{M.~S.} \bibnamefont{Safronova}}
  \bibnamefont{and} \bibinfo{author}{\bibfnamefont{U.~I.}
  \bibnamefont{Safronova}}, \bibinfo{journal}{Phys.\ Rev.\ A}
  \textbf{\bibinfo{volume}{83}}, \bibinfo{pages}{012503}
  (\bibinfo{year}{2011}).

\bibitem[{\citenamefont{Kotochigova and Tupitsyn}(1987)}]{KT87}
\bibinfo{author}{\bibfnamefont{S.~A.} \bibnamefont{Kotochigova}}
  \bibnamefont{and} \bibinfo{author}{\bibfnamefont{I.~I.}
  \bibnamefont{Tupitsyn}}, \bibinfo{journal}{J. Phys. B}
  \textbf{\bibinfo{volume}{20}}, \bibinfo{pages}{4759} (\bibinfo{year}{1987}).

\bibitem[{\citenamefont{Kozlov et~al.}(2015)\citenamefont{Kozlov, Porsev,
  Safronova, and Tupitsyn}}]{MBPT}
\bibinfo{author}{\bibfnamefont{M.~G.} \bibnamefont{Kozlov}},
  \bibinfo{author}{\bibfnamefont{S.~G.} \bibnamefont{Porsev}},
  \bibinfo{author}{\bibfnamefont{M.~S.} \bibnamefont{Safronova}},
  \bibnamefont{and} \bibinfo{author}{\bibfnamefont{I.~I.}
  \bibnamefont{Tupitsyn}}, \bibinfo{journal}{Comp. Phys. Comm.}
  \textbf{\bibinfo{volume}{195}}, \bibinfo{pages}{199} (\bibinfo{year}{2015}).

\bibitem[{\citenamefont{Dzuba et~al.}(1996)\citenamefont{Dzuba, Flambaum, and
  Kozlov}}]{DzuFlaKoz96}
\bibinfo{author}{\bibfnamefont{V.~A.} \bibnamefont{Dzuba}},
  \bibinfo{author}{\bibfnamefont{V.~V.} \bibnamefont{Flambaum}},
  \bibnamefont{and} \bibinfo{author}{\bibfnamefont{M.~G.}
  \bibnamefont{Kozlov}}, \bibinfo{journal}{Phys.\ Rev.\ A}
  \textbf{\bibinfo{volume}{54}}, \bibinfo{pages}{3948} (\bibinfo{year}{1996}).

\bibitem[{\citenamefont{Mann and Johnson}(1971)}]{ManJoh71}
\bibinfo{author}{\bibfnamefont{J.~B.} \bibnamefont{Mann}} \bibnamefont{and}
  \bibinfo{author}{\bibfnamefont{W.~R.} \bibnamefont{Johnson}},
  \bibinfo{journal}{Phys. Rev. A} \textbf{\bibinfo{volume}{4}},
  \bibinfo{pages}{41} (\bibinfo{year}{1971}).

\end{thebibliography}

\end{document}